\documentclass[aps,pre,11pt,nofootinbib,bibnotes,byrevtex,showkeys,a4paper,longbibliography]{revtex4-1}
\usepackage{microtype}
\usepackage{graphics}
\usepackage{amsthm}
\usepackage{amsmath}
\usepackage{amssymb}

\usepackage[colorlinks,linkcolor=blue,urlcolor=blue,citecolor=blue]{hyperref}

\newcommand{\be}{\begin{equation}}
\newcommand{\ee}{\end{equation}}
\newcommand{\om}{\omega}

\newcommand{\ra}{\rightarrow}

\newcommand{\reals}{\mathbb{R}}

\newcommand{\cE}{\mathcal{E}}
\newcommand{\cM}{\mathcal{M}}
\newcommand{\cA}{\mathcal{A}}
\newcommand{\cB}{\mathcal{B}}
\newcommand{\cF}{\mathcal{F}}
\newcommand{\cL}{\mathcal{L}}
\renewcommand{\Pr}{\mathbb{P}}
\newcommand{\Qr}{\mathbb{Q}}
\newcommand{\Ex}{\mathbb{E}}
\renewcommand{\dag}{\dagger}
\newcommand{\Fhat}{\hat{F}}
\newcommand{\micro}{\textrm{micro}}
\newcommand{\cano}{\textrm{cano}}
\newcommand{\id}{1\!\! 1}
\DeclareMathOperator{\tr}{tr}
\newcommand{\procequiv}{\cong}
\newcommand{\invrho}{\rho_\text{inv}}
\newcommand{\invmu}{\mu_\text{inv}}

\newtheoremstyle{myplain}
{5pt}			
{5pt}			
{\itshape}	
{}			
{\bfseries}		
{.}			
{.5em}		
{\thmname{#1}\thmnumber{ #2}\thmnote{~{(#3)}}}

\theoremstyle{myplain}
\newtheorem{theorem}{Theorem}

\newtheoremstyle{myplainex}
{5pt}			
{5pt}			
{\normalsize}		
{}			
{\bfseries}		
{.}			
{.5em}		
{\thmname{#1}\thmnumber{ #2}\thmnote{~{(#3)}}}

\theoremstyle{myplainex}
\newtheorem{example}[theorem]{Example}

\begin{document}

\title{Nonequilibrium Markov processes conditioned on large deviations}

\author{Rapha\"el Chetrite}
\affiliation{Laboratoire J.~A. Dieudonn\'e, UMR CNRS 7351, Universit\'e de Nice Sophia Antipolis, Nice 06108, France}

\author{Hugo Touchette}
\affiliation{National Institute for Theoretical Physics (NITheP), Stellenbosch 7600, South Africa}
\affiliation{Institute of Theoretical Physics, Stellenbosch University, Stellenbosch 7600, South Africa}

\date{\today}

\begin{abstract}
We consider the problem of conditioning a Markov process on a rare event and of representing this conditioned process by a conditioning-free process, called the effective or driven process. The basic assumption is that the rare event used in the conditioning is a large deviation-type event, characterized by a convex rate function. Under this assumption, we construct the driven process via a generalization of Doob's $h$-transform, used in the context of bridge processes, and show that this process is equivalent to the conditioned process in the long-time limit. The notion of equivalence that we consider is based on the logarithmic equivalence of path measures and implies that the two processes have the same typical states. In constructing the driven process, we also prove equivalence with the so-called exponential tilting of the Markov process, often used with importance sampling to simulate rare events and giving rise, from the point of view of statistical mechanics, to a nonequilibrium version of the canonical ensemble. Other links between our results and the topics of bridge processes, quasi-stationary distributions, stochastic control, and conditional limit theorems are mentioned.
\end{abstract}



\maketitle

\tableofcontents
\newpage

\section{Introduction}
\label{sec:intro}

We treat in this paper the problem of conditioning a Markov process $X_{t}$ on a rare event $\cA_T$ defined on the time interval $[0,T]$, and of representing this conditioned Markov process in terms of a conditioning-free Markov process $Y_t$, called the \emph{effective} or \emph{driven process}, having the same typical states as the conditioned process in the stationary limit $T\ra\infty$. More abstractly, this means that we are looking for a Markov process $Y_t$ such that
\be
X_t|\cA_T \procequiv Y_t,
\label{eqequiv1}
\ee
where $X_t|\cA_T$ stands for the conditioned process and $\procequiv$ is an asymptotic notion of process equivalence, related to the equivalence of ensembles in statistical physics, which we will come to define in a precise way below. Under some conditions on $X_t$, and for a certain class of large deviation-type events $\cA_T$, we will show that $Y_t$ exists and is unique, and will construct its generator explicitly.

This problem can be considered as a generalization of Doob's work on Markov conditioning~\cite{doob1957,doob1984} and also finds its source, from a more applied perspective, in many fundamental and seemingly unrelated problems of probability theory, stochastic simulations, optimal control theory, and nonequilibrium statistical mechanics. These are briefly discussed next to set the context of our work:

\medskip

\textbf{Conditioned Markov processes:} Doob was the first historically to consider conditioning of Markov processes, starting with the Wiener process conditioned on leaving the interval $[0,\ell]$ at the boundary $\{\ell\}$ \cite{doob1957,doob1984}. In solving this problem, he introduced a transformation of the Wiener process, now referred to as \emph{Doob's $h$-transform}, which was later adapted under the same name to deal with other conditionings of stochastic processes, including the Brownian bridge \cite{rogers2000}, Gaussian bridges \cite{baudoin2002,gasbarra2007,sottinen2014}, and the Schr\"odinger bridge \cite{schrodinger1931,schrodinger1932,jamison1975,zambrini1986,aebi1996}, obtained by conditioning a process on reaching a certain target distribution in time as opposed to a target point. Doob's transform also appears prominently in the theory of quasi-stationary distributions~\cite{darroch1965,darroch1967,villemonais2012,collet2014,doorn2013}, which describes in the simplest case the conditioning of a process never to reach an absorbing state.

We discuss some of these historical examples in Sec.~\ref{sec:gdt} to explain how Doob's original transform relates to the large deviation conditioning considered here. Following this section, we will see that the construction of the driven process $Y_t$ also gives rise to a process transformation, which is however different from Doob's transform because of the time-integrated character of the conditioning $\cA_T$ considered.

\medskip

\textbf{Gibbs conditioning and conditional limit theorems:} Let $X_1,\ldots,X_n$ be a sequence of independent and identically distributed random variables with common distribution $P(x)$ and let $S_n$ denote their sample mean:
\be
S_n=\frac{1}{n}\sum_{i=1}^n X_i.
\ee
A conditional limit theorem for this sequence refers to the distribution of $X_1$ obtained in the limit $n\ra\infty$ when the whole sequence $X_1,\ldots,X_n$ is conditioned on $S_n$ being in a certain interval or on $S_n$ assuming a certain value. In the latter case, it is known that, under some conditions on $P(x)$, 
\be
\lim_{n\ra\infty} P\{X_1=x|S_n=s\} =\frac{P(x)\, e^{kx}}{W(k)}\equiv P_k(x),
\ee
where $k$ is a real parameter related to the conditioning value $s$ and $W(k)$ is the generating function of $P(x)$ normalizing the so-called \emph{exponentially tilted} distribution $P_k(x)$; see \cite{vasicek1980,campenhout1981,cover1991,csiszar1984} for details. This asymptotic conditioning of a sequence of random variables is sometimes referred to as \emph{Gibbs conditioning} \cite{dembo1998} because of its similarity with the construction of the microcanonical ensemble of statistical mechanics, further discussed below. Other limit theorems can be obtained by considering sub-sequences of $X_1,\ldots,X_n$ instead of $X_1$, as above (see \cite{diaconis1987,diaconis1988,dembo1996}), or by assuming that the $X_i$'s form a Markov chain instead of being independent \cite{csiszar1987,borkar2004}.

This paper came partly as an attempt to generalize these results to general Markov processes and, in particular, to continuous-time processes. The essential step needed to arrive at these results is the derivation of the driven process; the conditional limit theorems that follow from this process will be discussed in a future publication.

\medskip

\textbf{Rare event simulations:} Many numerical methods used for determining rare event probabilities are based on the idea of importance sampling, whereby the underlying distribution $P$ of a random variable or process is modified to a target distribution $Q$ putting more weight on the rare events to be sampled \cite{bucklew2004}. A particularly useful and general distribution commonly used in this context is the exponentially tilted distribution $P_k$ mentioned earlier, which is also known as the \emph{exponential family} or \emph{Esscher transform} of $P$ \cite{feller1970}. Such a distribution can be generalized to sequences of random variables, as well as paths of stochastic processes (as a path measure), and corresponds, from the point of view of statistical mechanics, to the probability distribution defining the canonical ensemble, which describes thermodynamic systems coupled to a heat bath with inverse temperature $\beta =-k$.

This link with statistical mechanical ensembles is discussed in more detail below. For the conditioning problem treated here, we make contact with $P_k$ by using this distribution as an intermediate step to construct the driven process $Y_t$, as explained in Secs.~\ref{sec:tiltedproc} and \ref{sec:emp}. An interesting by-product of this construction is that we can interpret $Y_t$ as a modified Markov process that asymptotically realizes, in a sense to be made precise below, the exponential tilting of  $X_t$.

A further link with rare event sampling is established in that the semi-group or propagator of $Y_t$ is deeply related to Feynman--Kac functionals, which underlie cloning \cite{giardina2006,lecomte2007a,tailleur2009} and genealogical \cite{moral2004} methods also used for sampling rare events. In fact, we will see in Sec.~\ref{sec:modobs} that the driven process $Y_t$ is essentially a normalized version of a non-conservative process, whose generator is the so-called tilted generator of large deviations and whose dominant eigenvalue (when it exists) is the so-called scaled cumulant generating function -- the main quantity obtained by cloning methods \cite{giardina2006,lecomte2007a,tailleur2009}.

\medskip

\textbf{Stochastic control and large deviations:} The generalization of Doob's transform that we will discuss in Sec.~\ref{sec:gdt} has been considered by Fleming and Sheu in their work on control representations of Feynman--Kac-type partial differential equations (PDEs) \cite{fleming1978,fleming1982,fleming1985,fleming1985b}. The problem here is to consider a linear operator of the form $L+V(x)$, where $L$ is the generator of a Markov process, and to provide a stochastic representation of the solution $\phi(x,t)$ of the backward PDE
\be
\frac{\partial \phi}{\partial t}+(L+V)\phi=0,\quad t\leq T 
\ee
with final condition $\phi(x,T)=\Phi(x)$. The Feynman--Kac formula \cite{kac1949,stroock1979,revuz1999} provides, as is well known, a stochastic representation of $\phi(x,t)$ in terms of the expectation
\be
\phi(x,t)=\Ex[\Phi(X_T) e^{\int_t^T V(X_s) ds }|X_t=x].
\label{eq:scgfc}
\ee
The idea of Fleming and Sheu is to consider, instead of $\phi$, the \emph{logarithm} or \emph{Hopf--Cole transform} $I=-\ln \phi$, which solves the Hamilton--Jacobi-like PDE,
\be
\frac{\partial I}{\partial t}+(HI)-V(x)=0,
\label{eq:hj1}
\ee
where $(HI)=-e^{I}(Le^{-I})$, and to find a controlled process $X_t^u$ with generator $L^u$, so as to rewrite (\ref{eq:hj1}) as a dynamic programming equation:
\be
\frac{\partial I}{\partial t}+\min_{u}\{(L^uI)(x)+k_V(x,u)\}=0,
\label{eq:bhj}
\ee
where $k_V(x,u)$ is some cost function that depends on $V$, the system's state, and the controller's state. In this form, they show that $I$ represents the value function of the control problem, involving a Lagrangian dual to the Hamiltonian $H$; see \cite{fleming2006} for a more detailed description.

These results have been applied by Fleming and his collaborators to give control representations of various distributions related to exit problems \cite{fleming1978,fleming1982,fleming1985,fleming1985b}, dominant eigenvalues of linear operators \cite{sheu1984,fleming1987,fleming1997}, and optimal solutions of sensitive risk problems \cite{fleming1992b,fleming1995,dupuis1997b}, which aim at minimizing functionals having the exponential form of (\ref{eq:scgfc}). What is interesting in all these problems is that the generator $L^u$ of the optimally-controlled process is given by a Doob transform similar to the one we use to construct the conditioned process. In their work, Fleming \textit{et al.}\ do not interpret this transformation as a conditioning, but as an optimal change of measure between the controlled and reference processes. Such a change of measure has also been studied in physics more recently by Nemoto and Sasa \cite{nemoto2011,nemoto2011b,nemoto2014}. We will discuss these links in more detail in a future publication.

\medskip

\textbf{Fluctuation paths and fluctuation dynamics:} It is well known that rare transitions in dynamical systems perturbed by a small noise are effected by special trajectories known as reaction paths, fluctuation paths, most probable paths or instantons; see \cite{touchette2009} for a review. These paths are described mathematically by the Freidlin--Wentzell theory of large deviations \cite{freidlin1984}, and are fundamental for characterizing many noise-activated (escape-type) processes arising in chemical reactions, biological processes, magnetic systems, and glassy systems \cite{gardiner1985,kampen1992,risken1996}.

The  concept of fluctuation path is specific to the low-noise limit: for processes with arbitrary random perturbations, there is generally not a single fluctuation path giving rise to a rare event, but many different fluctuation paths leading to the same event, giving rise to what we call a \emph{fluctuation dynamics}. The driven process that we construct in this paper is a specific example of such a fluctuation dynamics: it describes the effective dynamics of $X_t$ as this process is seen to fluctuate away from its typical behavior to `reach' the event $\cA_T$. Consequently, it can be used to simulate or sample this fluctuation in an efficient way, bringing yet another connection with rare event simulations. This will be made clearer as we come to define this process in Sec.~\ref{sec:emp}. 

\medskip

\textbf{Statistical ensembles for nonequilibrium systems:} The problem of defining or extending statistical ensembles, such as the microcanonical and canonical ensembles, to nonequilibrium systems has a long history in physics. It was revived recently by Evans \cite{evans2004,evans2005a,evans2010}, who proposed deriving the transition rates of a system driven by external forces in a stationary nonequilibrium state by conditioning the transition rates of the same system when it is not driven, that is, when it is in an equilibrium state with transition rates satisfying detailed balance. Underlying this proposal is the interesting idea that nonequilibrium systems driven in steady states could be seen as equilibrium systems in which the driving is effected by a conditioning. This means, for example, that a driven nonequilibrium system having a given stationary particle current could be thought of, physically, as being equivalent to a non-driven equilibrium system in which this current appears as a fluctuation.

The validity of this idea needs to be tested using examples of driven physical systems for which nonequilibrium stationary solutions can be obtained explicitly and be compared with conditionings of their equilibrium solutions. Our goal here is not to provide such a test, but to formalize the problem in a clear, mathematical way as a Markov conditioning problem based on large deviations. This leads us to define in a natural way a nonequilibrium generalization of the microcanonical ensemble for trajectories or paths of Markov processes, as well as a nonequilibrium version of the canonical ensemble, which is a path version of the exponentially tilted measure $P_k$. 

The latter ensemble has been used recently with transition path sampling \cite{dellago2003,dellago2006,dellago2009,vanden-eijnden2006b} to simulate rare trajectories of nonequilibrium systems associated with glassy phases and dynamical phase transitions; see \cite{chandler2010} for a recent review. In this context, the exponentially tilted distribution $P_k$ is referred to as the \emph{biased}, \emph{tilted} or \emph{$s$-ensemble}, the last name stemming from the fact that the symbol $s$ is used instead of $k$ \cite{hedges2009,jack2010b,lecomte2005,lecomte2007,chandler2010}. These simulations follow exactly the idea of importance sampling mentioned earlier: they re-weight the statistics of the trajectories or paths of a system in an exponential way so as to reveal, in a typical way, trajectories responsible for certain states or phases that are atypical in the original system. In Sec.~\ref{sec:emp}, we will give conditions that ensure that this exponential re-weighting is equivalent to a large deviation conditioning -- in other words, we will give conditions ensuring that the path canonical ensemble is equivalent to the path microcanonical ensemble. 

The connection with the driven process is established from this equivalence by showing that the canonical ensemble can be realized by a Markov process in the long-time limit. Some results on this canonical--Markov connection were obtained by Jack and Sollich \cite{jack2010b} for a class of jump processes and by Garrahan and Lesanovsky~\cite{garrahan2010}  for dissipative quantum systems (see also \cite{garrahan2011,ates2012,genway2012,hickey2012}). Here, we extend these results to general Markov processes, including diffusions, and relate them explicitly to the conditioning problem.

\medskip

These connections and applications will not be discussed further in the paper, but should hopefully become clearer as we define the driven process and study its properties in the next sections. The main steps leading to this process are summarized in \cite{chetrite2013}; here, we provide the full derivation of this process and discuss, as mentioned, its link with Doob's results. We also discuss new results related to constraints satisfied by the driven process, as well as special cases of these results for Markov chains, jump processes, and pure diffusions.

The plan of the paper is as follows. In Sec.~\ref{sec:modobs}, we define the class of general Markov processes and conditioning events (or observables) that we consider, and introduce various mathematical concepts (Markov semi-groups, Markov generators, path measures) used throughout the paper. We also define in that section the path versions of the microcanonical and canonical ensembles, corresponding respectively to the conditioning and exponential tilting of $X_t$, and introduce all the elements of large deviation theory needed to define and study our class of rare event conditioning. We then proceed to construct the driven process $Y_t$ and prove its equivalence with the conditioned process $X_t|\cA_T$ in three steps. Firstly, we construct in Sec.~\ref{sec:tiltedproc} a non-conservative process from which various spectral elements, related to the large deviation conditioning, are obtained. Secondly, we study in Sec.~\ref{sec:gdt} the generalization of Doob's transform needed to construct $Y_t$, and show how it relates to the original transform considered by Doob. Thirdly, we use the generalized transform to define in Sec.~\ref{sec:emp} the driven process proper, and show that it is equivalent to the conditioned process by appealing to general results about ensemble equivalence. 

Our main results are contained in Sec.~\ref{sec:emp}. Their novelty, compared to previous works, resides in the fact that we treat the equivalence of the driven and conditioned processes explicitly via path versions of the canonical and microcanonical ensembles, derive precise conditions for this equivalence to hold, and express all of our results in the general language of Markov generators, which can be used to describe jump processes, diffusions, or mixed processes, depending on the physical application considered. New properties of the driven process, including constraint rules satisfied by its transition rates or generator, are also discussed in that section. Section~\ref{sec:apps} finally presents some applications of our results for diffusions, to show how the driven process is obtained in practice, and for absorbing Markov chains, to make a connection with quasi-stationary distributions. The specialization of our results to Markov chains is summarized in the Appendices, which also collect various technical steps needed for proving our results.

\newpage
\section{Notations and definitions}
\label{sec:modobs}

We define in this section the class of Markov processes and observables of these processes that we use to define the rare event conditioning problem. Markov processes are widely used as models of stochastic systems, for example, in the context of financial time series \cite{karatzas1998}, biological processes \cite{berg1993}, and chemical reactions \cite{gardiner1985,kampen1992,risken1996}. In physics, they are also used as a general framework for modeling systems driven in nonequilibrium steady states by noise and external forces \cite{gardiner1985,kampen1992,risken1996}, such as interacting particle systems coupled to different particle and energy reservoirs, which have been studied actively in the mathematics and physics literature recently \cite{spohn1991,kipnis1999,liggett2004,derrida2007,bertini2007}. For general introductions to Markov processes and their applications in physics, see \cite{gardiner1985,kampen1992,risken1996,jacobs2010,krapivsky2010,nelson1967}; for references on the mathematics of these processes, see \cite{stroock1979,anderson1991,revuz1999,rogers2000,chung2005}.

\subsection{Homogeneous Markov processes}
\label{sub:markov}

We consider a homogeneous continuous-time Markov process $X_{t}$, with $t\in \reals_+$, taking values in some space $\cE$, which, for concreteness, is assumed to be $\reals^d$ or a counting space.\footnote{In probability theory, $\cE$ is most often taken to be a so-called Polish (metric, separable and complete) topological space.} The dynamics of $X_t$ is described by a transition kernel $P_{t}(x,dy)$ giving the conditional probability that $X_{t+t'}\in dy$ given that $X_{t'}=x$ with $t\geq 0$. This kernel satisfies the Chapmann--Kolmogorov equation
\be
\int_{\cE}P_{t'}(x,dy)P_{t}(y,dz)\,=\, P_{t'+t}(x,dz)
\label{CK}
\ee
for all $(x,z)\in\cE^2$, and is homogeneous in the sense that it depends only on the time difference $t$ between $X_{t+t'}$ and $X_{t'}$. Here and in the following, $dy$ stands for the Lebesgue measure or the counting measure, depending on $\cE$. 

To ensure that $X_t$ is well behaved, we assume that it admits c\`adl\`ag\footnote{From the French `continue \`a droite, limite \`a gauche': right continuous with left limit.} paths as a function of time for every initial condition $X_0=x\in\cE$. We further assume that
\be
\int_{\cE}P_{t}(x,dy)=1,
\ee
so that the probability is conserved at all times. This property is also expressed in the literature by saying that $X_t$ is \emph{conservative}, \emph{honest}, \emph{stochastically complete} or \emph{strictly Markovian}, and only means physically that there is no killing or creation of probability. Although $X_t$ is assumed to be conservative, we will introduce later a non-conservative process as an intermediate mathematical step to construct the driven process. In what follows, it will be clear when we are dealing with a conservative or a non-conservative process. Moreover, it should be clear that the word `conservative' is not intended here to mean that energy is conserved.

Mathematically, the transition kernel can be thought of as a positive linear operator\footnote{This operator is positive in the Perron--Frobenius sense, that is, $(P_t f)\geq 0$ for all $f\geq 0$.} acting on the space of bounded measurable functions $f$ on $\cE$ according to
\be
(P_{t}f)(x)\equiv\int_{\cE}P_{t}(x,dy)f(y)\equiv \Ex_x[f(X_t)]
\ee
for all $x\in\cE$, where $\Ex_x[\cdot]$ denotes the expectation with initial condition $X_0=x$. In many cases, it is more convenient to give a local specification of the action of $P_t$ via its \emph{generator} $L$ according to
\be
\partial_t \Ex_x[f(X_t)]=\Ex_x[(Lf)(X_t)],
\ee
where $(Lf)$ denotes the application of $L$ on $f$. Formally, this is equivalent to the representation
\be
P_{t}=e^{tL},
\label{gen}
\ee
and the forward and backward Kolmogorov equation, given by
\be
\partial_{t}P_{t}=P_{t}L=LP_{t},\qquad P_{0}=I,
\label{FK}
\ee
where $I$ is the identity operator. For $P_t(x,dy)$ to be conservative, the generator must obey the relation $(L1)=0$, where $1$ is the constant function equal to $1$ on $\cE$.

In the following, we will appeal to a different characterization of $X_t$ based on the path probability measure $d\Pr_{L,\mu_0,T}(\om)$ representing, roughly speaking, the probability of a trajectory or sample path $\{X_t(\om)\}_{t=0}^T$ over the time interval $[0,T]$, with $X_0(\om)$ chosen according to the initial measure $\mu_0$. Technically, the space of such paths is defined as the so-called \emph{Skorohod space} $D([0,T],\cE)$ of c\`adl\`ag functions on $\cE$, while $d\Pr_{L,\mu_0,T}(\om)$ is defined in terms of expectations having the form
\be
\Ex_{\mu_0}[C] =\int C(\om)\, d\Pr_{L,\mu_0,T}(\om),
\label{eq:traj}
\ee
where $C$ is any bounded measurable functional of the path $\{X_t(\om)\}_{t=0}^T$, and $\Ex_{\mu_0}$ now denotes the expectation with initial measure $\mu_0$. As usual, this expectation can be simplified to completely characterize $\Pr_{L,\mu_0,T}$ by considering so-called \emph{cylinder functions},
\be
C(\om)=C\left(X_{0}(\om),X_{t_{1}}(\om),...,X_{t_{n-1}}(\om),X_{T}(\om)\right),
\label{eq:cy}
\ee
involving $X_t$ over a finite sequence of times $0\leq t_{1}\leq t_{2}\leq\cdots\leq t_{n-1}\leq T$ instead of the whole interval $[0,T]$. At this level, the path probability measure becomes a joint probability distribution over these times, given in terms of $L$ by
\be
\Pr_{L,\mu_0,T}(dx_0,\ldots,dx_n)=\mu_0(dx_0)\; e^{t_1L }(x_0,dx_1)\; e^{(t_2-t_1)L}(x_{1},dx_{2})\cdots e^{(T-t_{n-1})L}(x_{n-1},dx_{n}),
\label{eq:cyl}
\ee
where the exponentials refer to the operator of (\ref{gen}).
 
One important probability measure obtained from the path measure is the marginal $\mu_t$ of $X_t$, associated with the single-time cylinder expectation,
\be
\Ex_{\mu_0}\left[C(X_{t})\right]=\int_{\cE}C(y)\, \mu_{t}(dy).
\label{eq:1p}
\ee
This measure is also obtained by `propagating' the initial measure $\mu_0$ according to (\ref{eq:cyl}):
\be
\mu_{t}(dy)=\int_{\cE}\mu_0(dx_0)\; e^{tL}(x_0,dy).
\ee
It then follows from the Kolmogorov equation (\ref{FK}) that
\be
\partial_{t}\mu_{t}(x)=(L^{\dag}\mu_{t})(x),
\label{eqm}
\ee
where $L^{\dag}$ is the formal adjoint of $L$ with respect to the Lebesgue or counting measure. In physics, this equation is referred to as the \emph{Master equation} in the context of jump processes or the \emph{Fokker--Planck equation} in the context of diffusions.

The time-independent probability measure $\invmu$ satisfying 
\be
(L^{\dag}\invmu)=0
\label{INV}
\ee
is called the \emph{invariant measure} when it exists. Furthermore, one says that the process $X_t$ is an \emph{equilibrium process} (with respect to $\invmu$) if its transition kernel satisfies the \emph{detailed balance condition},
\be
\invmu(dx)P_{t}(x,dy)=\invmu(dy)P_t(y,dx)
\label{DB}
\ee
for all $(x,y)\in\cE^{2}$. In the case where $\invmu$ has the density $\invrho(x)\equiv\invmu(dx)/dx$ with respect to the Lebesgue or counting measure, this condition
can be expressed as the following operator identity for the generator:
\be
\invrho L\invrho^{-1}=L^{\dag},
\label{DB'}
\ee
which is equivalent to saying that $L$ is self-adjoint with respect to $\invmu$. If the process $X_t$ does not satisfy this condition, then it is referred to in physics as a \emph{nonequilibrium} Markov process. Here, we follow this terminology and consider both equilibrium and nonequilibrium processes.

\subsection{Pure jump processes and diffusions}

Two important types of Markov processes will be used in this paper to illustrate our results, namely, pure jump processes and diffusions. In continuous time and continuous space, all Markov processes consist of a superposition of these two processes, combined possibly with deterministic motion \cite{rogers2000,applebaum2009,sato1999}. The case of discrete-time Markov chains is discussed in Appendix~\ref{app:mc}.

A homogeneous Markov process $X_t$ is a pure jump process if the probability that $X_t$ undergoes one jump during the time interval $[t,t+dt]$ is proportional to $dt$.\footnote{In a countable space, one can show that all Markov processes with right continuous paths are of this type, a property which is not true in a general space \cite{applebaum2009,sato1999}.} To describe these jumps, it is usual to introduce the bounded intensity or \emph{escape rate} function $\lambda(x)$, such that $\lambda(x)dt+o(dt)$ is the probability that $X_{t}$ undergoes a jump during $\left[t,t+dt\right]$ starting from the state $X_{t}=x$. When a jump occurs, $X(t+dt)$ is then distributed with the kernel $T(x,dy)$, so that the overall transition rate is 
\be
W(x,dy)\equiv\lambda(x)\, T(x,dy)
\label{eq:gps}
\ee
for $(x,y)\in\cE^2$. Over a time interval $[0,T]$, the path of such a process can thus be represented by the sequence of visited states in $\cE$, together with the sequence of waiting times in those states, so that the space of paths is $[\cE\times(0,\infty)]^{\mathbb{N}}$.

Under some regularity conditions (see \cite{revuz1999,anderson1991}), one can show that this process possesses a generator, given by
\be
(Lf)(x)=\int_{\cE}W(x,dy)[f(y)-f(x)]
\label{eq:genps}
\ee
for all bounded, measurable function $f$ defined on $\cE$ and all $x\in\cE$. In terms of transition rates, the condition of detailed balance with respect to some invariant measure $\invmu$ is expressed as
\be
\invmu(dx)W(x,dy)=\invmu(dy)W(y,dx)
\label{eq:BDS}
\ee
for all $(x,y)\in\cE^2$. 

Pure diffusions driven by Gaussian white noise have, contrary to jump processes, continuous sample paths and are best described not in terms of transition rates, but in terms of stochastic differential equations (SDEs). For $\cE=\reals^d$, these have the general form:
\begin{equation}
dX_t=F(X_t)dt+\sum_{\alpha} \sigma_\alpha(X_t) \circ dW_{\alpha}(t),
\label{eq:diff}
\end{equation}
where $F$ and $\sigma_\alpha$ are smooth vector fields on $\reals^d$, called respectively the \emph{drift} and \emph{diffusion coefficient}, and $W_\alpha$ are independent Wiener processes (in arbitrary number, so that the range of $\alpha$ is left unspecified). The symbol $\circ$ denotes the Stratonovich (midpoint) convention used for interpreting the SDE; the It\=o convention can also be used with the appropriate changes.

In the Stratonovich convention, the explicit form of the generator is
\be
L=F\cdot\nabla+\frac{1}{2}\sum_{\alpha}(\sigma_\alpha \cdot \nabla)^{2}=\Fhat \cdot \nabla+\frac{1}{2}\nabla D \nabla,
\label{eq:gpd}
\ee
where
\be
\Fhat(x)=F(x)-\frac{1}{2}\sum_{\alpha} (\nabla \cdot \sigma_\alpha)(x)\, \sigma_\alpha(x)
\label{eq:vhat}
\ee
is the so-called modified drift and 
\be
D^{ij}(x)=\sum_{\alpha}\sigma_{\alpha}^{i}(x)\sigma_{\alpha}^{j}(x)
\label{eq:diffel}
\ee
is the covariance matrix involving the components of $\sigma_\alpha$. The notation $\nabla D\nabla$ in (\ref{eq:gpd}) is a shorthand for the operator
\be
\nabla D\nabla =\sum_{i,j} \frac{\partial}{\partial x_i} D^{ij}(x) \frac{\partial}{\partial x_j},
\label{eq:dp}
\ee
which is also sometimes expressed as $\nabla\cdot (D\nabla)$ or in terms of a matrix trace as $\tr D\nabla^2$. With these notations, the condition of detailed balance for an invariant measure $\invmu(dx)$, with density $\invrho(x)$ with respect to the Lebesgue measure,\footnote{This density exists, for example, when the conditions of Hormander's Theorem are satisfied \cite{hormander1967,malliavin1976}.} is equivalent to
\be
\Fhat=\frac{D}{2}\nabla\ln\invrho.
\label{eq:BDD}
\ee
Similar results can be obtained for the It\=o interpretation. Obviously, the need to distinguish the two interpretations arises only if the diffusion fields $\sigma_\alpha$ depend on $x\in\cE$. If these fields are constant, then the Stratonovich and It\=o interpretations yield the same results with $\Fhat=F$ and $\nabla D\nabla=D\nabla^2$.

\subsection{Conditioning observables}
\label{sub:obs}

Having defined the class of stochastic processes of interest, we now define the class of events $\cA_T$ used to condition these processes. The idea is to consider a random variable or \emph{observable} $A_T$, taken to be a real function of the paths of $X_t$ over the time interval $[0,T]$, and to condition $X_t$ on a general measurable event of the form $\cA_{T}=\{A_T\in B\}$ with $B\subset\reals$. This means, more precisely, that we condition $X_t$ on the subset
\be
\cA_T=\{\om \in D([0,T],\cE): A_T(\om)\in B\}
\ee
of sample paths satisfying the constraint that $A_T\in B$. In the following, we will consider the smallest event possible, $\{A_T= a\}$, representing the set of paths for which $A_T$ is contained in the infinitesimal interval $[a,a+da]$ or, more formally, the set of paths such that $A_T(\om)=a$. General conditionings of the form $\{A_T\in B\}$ can be treated by integration over $a$. We then write $X_t|A_T = a$ to mean that the process $X_t$ is conditioned on the basic event $\{A_T=a\}$. Formally, we can also study this conditioning by considering path probability densities instead of path measures, as done in \cite{chetrite2013}. 

Mathematically, the observable $A_T$ is assumed to be non-anticipating, in the sense that it is adapted to the natural ($\sigma$-algebra) filtration $\cF_T=\sigma\{X_t(\om):0\leq t\leq T\}$  of the process up to time $T$. Physically, we also demand that $A_T$ depend only on $X_t$ and its transitions or displacements. For a pure jump process, this means that we consider a general observable of the form
\be
A_{T}=\frac{1}{T}\int_{0}^{T} f(X_{t})dt + \frac{1}{T} \sum_{0\leq t\leq T:\Delta X_{t}\neq0} g(X_{t^{-}},X_{t^{+}}),
\label{eq:osp-1}
\ee
where $f:\cE\ra\reals$, $g:\cE^2\ra\reals$, and $X_{t^-}$ and $X_{t^+}$ denote, respectively, the state of $X_t$ before and after a jump at time $t$. The discrete sum over the jumps of the process is well defined, since we suppose that $X_t$ has a finite number of jumps in $[0,T]$ with probability one. 

The class of observables $A_T$ defined by $f$ and $g$ includes many random variables of mathematical interest, such as the number of jumps over $[0,T]$, obtained with $f=0$ and $g=1$, or the occupation time in some set $\Delta$, obtained with $f(x)=\id_\Delta(x)$ and $g=0$, with $\id_\Delta$ the characteristic function of the set $\Delta$. From a physical point of view, it also includes many interesting quantities, including the fluctuating entropy production \cite{lebowitz1999}, particle and energy currents \cite{derrida2007}, the so-called activity \cite{lecomte2005,lecomte2007,maes2008a,baiesi2009}, which is essentially the number of jumps, in addition to work- and heat-related quantities defined for systems in contact with heat reservoirs and driven by external forces \cite{jarzynski1997b,crooks1998}.

For a pure diffusion process $X_t\in\reals^d$, the appropriate generalization of the observable above is
\be
A_{T} = \frac{1}{T}\int_{0}^{T} f(X_{t}) dt +\frac{1}{T} \int_{0}^{T}\sum_{i=1}^{d}g^{i}(X_{t})\circ dX_{t}^{i},
\label{eq:opd-1}
\ee
where $f:\cE\ra\reals$, $g:\cE\ra\reals^d$, $\circ$ denotes as before the Stratonovich product, and $g^i$ and $X_t^i$ are the components of $g$ and $X_t$, respectively. This class of `diffusive' observables defined by the function $f$ and the vector field $g$ also includes many random variables of mathematical and physical interest, including occupation times, empirical distributions, empirical currents or flows, the fluctuating entropy production \cite{lebowitz1999}, as well as work and heat quantities \cite{sekimoto2010}. For example, the empirical density of $X_t$, which represents the fraction of time spent at $x$, is obtained formally by choosing $f(y)=\delta(y-x)$ and $g=0$, while the empirical current, recently considered in the physics literature \cite{maes2008,chernyak2009}, is defined, also formally, with $f=0$ and $g(y)=\delta(y-x)$. 

The consideration of diffusions and current-type observables of the form (\ref{eq:opd-1}) involving a stochastic integral is one of the main contributions of this paper, generalizing previous results obtained by Jack and Sollich \cite{jack2010b} for jump processes, Garrahan and Lesanovsky~\cite{garrahan2010} for dissipative quantum systems, and by Borkar \textit{et al.}~\cite{borkar2004,basu2008} for Markov chains. 

\subsection{Large deviation principle}

As mentioned in the introduction, the conditioning event $\cA_T$ must have the property of being atypical with respect to the measure of $X_t$, otherwise the conditioning should have no effect on this process in the asymptotic limit $T\ra\infty$. Here, we assume that $\{A_T= a\}$ is \emph{exponentially} rare with $T$ with respect to the measure $\Pr_{L,\mu_0,T}$ of  $X_t$, which means that we define this rare event as a \emph{large deviation event}. This exponential decay of probabilities applies to many systems and observables of physical and mathematical interest, and is defined in a precise way as follows. The random variable $A_T$ is said to satisfy a \emph{large deviation principle} (LDP) with respect to $\Pr_{L,\mu_0,T}$ if there exists a lower semi-continuous function $I$ such that
\be
\liminf_{T\ra\infty} -\frac{1}{T}\ln \Pr_{L,\mu_0,T}\{A_T\in C\}\geq \inf_{a\in C} I(a)
\ee
\emph{for any closed sets $C$ and}
\be
\limsup_{T\ra\infty} - \frac{1}{T}\ln \Pr_{L,\mu_0,T}\{A_T\in O\}\leq \inf_{a\in O} I(a)
\ee
\emph{for any open sets $O$} \cite{dembo1998,hollander2000,ellis1985}. The function $I$ is called the \emph{rate function}. 

The basic assumption of our work is that the function $I$ exists and is different from $0$ or $\infty$. If the process $X_t$ is ergodic, then an LDP for the class of observables $A_T$ defined above holds, at least formally, as these observables can be obtained by contraction of the so-called level 2.5 of large deviations concerned with the empirical density and empirical current. This level has been studied formally in \cite{maes2008,maes2008a,chernyak2009}, and rigorously for jump processes with finite space in \cite{fortelle2001} and countable space in \cite{bertini2012}. The observable $A_T$ can also satisfy an LDP if the process $X_t$ is not ergodic; in this case, however, the existence of the LDP must be proved on a process by process basis and may depend on the initial condition of the process considered.

Formally, the existence of the LDP is equivalent to assuming that
\be
\lim_{T\ra\infty}-\frac{1}{T}\ln \Pr_{L,\mu_0,T}\{A_T\in [a,a+da]\}=I(a),
\label{eqldp}
\ee
so that the measure $\Pr_{L,\mu_0,T}\{A_T\in [a,a+da]\}$ decays exponentially with $T$, as mentioned. The fact that this decay is in general not exactly, but only approximately exponential is often expressed by writing 
\be
\Pr_{L,\mu_0,T}\{A_T\in [a,a+da]\}\asymp e^{-TI(a)}\, da,
\ee
where the approximation $\asymp$ is defined according to the large deviation limit (\ref{eqldp}) \cite{ellis1985,touchette2009}. We will see in the next subsection that this exponential approximation, referred to in information theory as the \emph{logarithmic} equivalence \cite{cover1991}, sets a natural scale for defining two processes as being equivalent in the stationary limit $T\ra\infty$.
 
\subsection{Nonequilibrium path ensembles}
\label{sub:noneqens}

We now have all the notations needed to define our problem of large deviation conditioning. At the level of path measures, the conditioned process $X_t|A_T= a$ is defined by the path measure
\be
d\Pr_{a,\mu_0,T}^\micro (\om) \equiv d\Pr_{L,\mu_0,T}\{\om |A_T= a\},
\label{eq:micro}
\ee
which is a pathwise conditioning of the reference measure $\Pr_{L,\mu_0,T}$ of $X_t$ on the value $A_T=a$ after the time $T$. By Bayes's Theorem, this is equal to 
\be
d\Pr^\micro_{a,\mu_0,T} (\om)=\frac{d\Pr_{L,\mu_0,T}(d\om)}{\Pr_{L,\mu_0,T} \{A_T=a\}}\; \id_{[a,a+da]}\left(A_T(\om)\right),
\label{eq:microb}
\ee
where $\id_\Delta(x)$ is, as before, the indicator (or characteristic) function of the set $\Delta$. We refer to this measure as the \emph{path microcanonical ensemble} (superscript micro)  \cite{evans2004,evans2005a,evans2010} because it is effectively a path generalization of the microcanonical ensemble of equilibrium statistical mechanics, in which the microscopic configurations of a system are conditioned or constrained to have a certain energy value. This energy is here replaced by the general observable $A_T$.

Our goal for the rest of the paper is to show that the microcanonical measure can be expressed or realized in the limit $T\ra\infty$ by a conservative Markov process, called the \emph{driven process}. This process will be constructed, as mentioned in the introduction, indirectly via another path measure, known as the \emph{exponential tilting} of $d\Pr_{L,\mu_0,T}(\om)$:
\be
d\Pr^\cano_{k,\mu_0,T}(\om)\equiv\frac{e^{TkA_T(\om)}\, d\Pr_{L,\mu_0,T}(\om)}{\Ex_{\mu_0}[e^{kTA_T}]},
\label{eq:cano}
\ee
where $k\in\reals$. In mathematics, this measure is also referred to as a \emph{penalization} or a \emph{Feynman--Kac transform} of $\Pr_{L,\mu_0,T}$ \cite{roynette2009}, in addition to the names `exponential family' and `Essher transform' mentioned in the introduction. In physics, it is referred, as also mentioned,  to as the biased, twisted, or $s$-ensemble, the last name arising again because the letter $s$ is often used in place of $k$ \cite{hedges2009,jack2010b,lecomte2005,lecomte2007,chandler2010}. We use the name `canonical ensemble' (superscript cano) because this measure is a path generalization of the well-known canonical ensemble of equilibrium statistical. From this analogy, we can interpret $k$ as the analog of a (negative) inverse temperature and the normalization factor $\Ex_{\mu_0}[e^{kTA_T}]$ as the analog of the partition function. 

The plan for deriving the driven process is to define a process $Y_t$ via a generalization of Doob's transform and to show that its path measure is equivalent in the asymptotic limit to the path canonical ensemble. Following this result, we will then use established results of ensemble equivalence to show that the canonical path ensemble is equivalent to the microcanonical path ensemble, so as to finally obtain the result announced in (\ref{eqequiv1}). The notion of measure or process equivalence underlying these results, denoted by $\procequiv$ in (\ref{eqequiv1}), is defined next.

\subsection{Process equivalence}
\label{subsec:procequiv}

Let $\Pr_T$ and $\Qr_T$ be two path measures associated with a Markov process over the time interval $[0,T]$. Assume that $\Pr_T$ is absolutely continuous with respect to $\Qr_T$, so that the Radon--Nikodym derivative $d\Pr_T/d\Qr_T$ exists. We say that $\Pr_T$ and $\Qr_T$ are \emph{asymptotically equivalent} if 
\be
\lim_{T\ra\infty}\frac{1}{T}\ln \frac{d\Pr_T}{d\Qr_T}(\om)=0
\ee
almost everywhere with respect to both $\Pr_T$ and $\Qr_T$. In this case, we also say that the Markov process $X_t$ defined by $\Pr_T$ and the different Markov process $Y_t$ defined by $\Qr_T$ are asymptotically equivalent, and denote this property by $X_t\procequiv Y_t$ as in (\ref{eqequiv1}).

This notion of process equivalence can be interpreted in two ways. Mathematically, it implies that $\Pr_T$ and $\Qr_T$ are logarithmically equivalent for most paths, that is, 
\be
d\Pr_T(\om)\asymp d\Qr_T(\om)
\ee
for almost all $\om$ with respect to $\Pr_T$ or $\Qr_T$. This is a generalization of the so-called \emph{asymptotic equipartition property} of information theory \cite{cover1991}, which states that the probability of sequences generated by an ergodic discrete source is approximately (i.e., logarithmically) constant for almost all sequences \cite{cover1991}. Here, we have that, although $\Pr_T$ and $\Qr_T$ may be different measures, they are approximately equal in the limit $T\ra\infty$ for almost all paths with respect to these measures.

In a more concrete way, the asymptotic equivalence of $\Pr_T$ and $\Qr_T$ also implies that an observable satisfying LDPs with respect to these measures concentrate on the same values for both measures in the limit $T\ra\infty$. In other words, the two measures lead to the same typical or ergodic states of (dynamic) observables in the long-time limit. A more precise statement of this result based on the LDP will be given when we come to proving explicitly the equivalence of the driven and conditioned processes. For now, the only important point to keep in mind is that the typical properties of the two processes $X_t$ and $Y_t$ such that $X_t\procequiv Y_t$ are essentially the same. This is a useful notion of equivalence when considering nonequilibrium systems, which is a direct generalization of the notion of equivalence used for equilibrium systems \cite{lewis1994,lewis1995,touchette2014}. For the latter systems, typical values of (static) observables are simply called \emph{equilibrium states}.

\section{Non-conservative tilted process}
\label{sec:tiltedproc}

We discuss in this section the properties of a non-conservative process associated with the canonical path measure (\ref{eq:cano}). This process is important as it allows us to obtain a number of important quantities related to the large deviations of $A_T$, in addition to giving some clues as to how the driven process will be constructed.
 
\subsection{Definition}

We consider as before a Markov process $X_t$ with path measure $\Pr_{L,\mu_0,T}$ and an observable $A_T$ defined as in (\ref{eq:osp-1}) or (\ref{eq:opd-1}) according to the type (jump process or diffusion, respectively) of $X_t$. From the path measure of $X_t$, we define a new path measure by
\be
d\Pr_{\cL_k,\mu_0,T}(\om)\equiv d\Pr_{L,\mu_0,T}(\om)\, e^{kT A_T(\om)},
\label{eq:ptilt'}
\ee
which corresponds to the numerator of the canonical path ensemble $d\Pr_{k,\mu_0,T}^\cano$, defined in (\ref{eq:cano}). As suggested by the notation, the new measure $d\Pr_{\cL_k,\mu_0,T}$ defines a Markov process of generator $\cL_{k}$, which we call the \emph{non-conservative tilted process}. This process is Markovian in the sense that 
\be
\Ex_{\mu_0}[e^{kTA_T}C]=\int_{\cE^{n+1}}C(x_0,\ldots,x_{n})\, \mu_0(dx_0)\, e^{t_1\cL_k}(x_0,dx_{1})\cdots e^{(T-t_{n-1})\cL_k}(x_{n-1},dx_{n}),
\label{eq:ptilt}
\ee
for any cylinder functional $C$ (\ref{eq:cy}), and is non-conservative because $(\cL_k 1)\neq 0$ in general.

The class of observables defined by (\ref{eq:osp-1}) and (\ref{eq:opd-1}) can be characterized in the context of this result as the largest class of random variables for which the Markov property above holds. The proof of this property cannot be given for arbitrary Markov processes, but is relatively straightforward when considering jump processes or diffusions. In each case, the proof of (\ref{eq:ptilt}) and the form of the so-called \emph{tilted generator} $\cL_k$ follow by applying Girsanov's Theorem and the Feynman--Kac formula, as shown in Appendix~\ref{app:psp} for jump processes and Appendix~\ref{app:df} for diffusions. The result in the first case is 
\be
(\cL_k h)(x)=\left( \int_{\cE} W(x,dy) [e^{kg(x,y)}h(y)-h(x)] \right)+kf(x)h(x)
\label{eq:gtps}
\ee
for all function $h$ on $\cE$ and all $x\in\cE$, where $f$ and $g$ are defined as in (\ref{eq:osp-1}). This can be written more compactly as 
\be
\cL_k=We^{kg} -(W1) +kf,
\label{eq:gtps2}
\ee
where the first term  is understood as the Hadamard (component-wise) product $W(x,dy)e^{k g(x,y)}$ and $kf$ is a diagonal operator $k(x)f(x)\delta(x-y)$. In the case of diffusions, we obtain instead
\be
\cL_k=\Fhat\cdot (\nabla+kg)+\frac{1}{2}(\nabla+kg) D(\nabla +kg)+kf,
\label{eq:gtpd}
\ee
where $f$ and $g$ are the functions appearing in (\ref{eq:opd-1}), while $\Fhat$ and $D$ are defined as in (\ref{eq:vhat}) and (\ref{eq:diffel}), respectively. The double product involving $D$ is defined as in (\ref{eq:dp}).

\subsection{Spectral elements}
\label{sub:spetilt}

The operator $\cL_k$ defined in (\ref{eq:gtps}) or (\ref{eq:gtpd}) is a Perron--Frobenius operator or, more precisely, a \emph{Metzler operator} with negative `diagonal' part \cite{maccluer2000}. The extension of the Perron--Frobenius Theorem to infinite-dimensional, compact operators is ruled by the Krein--Rutman Theorem \cite{krein1950}. For differential elliptic operators having the form (\ref{eq:gtpd}), this theorem can be applied on compact and smooth domains with Dirichlet boundary conditions. 

We denote by $\Lambda_{k}$ the real dominant (or principal) eigenvalue of $\cL_k$ and by $r_k$ its associated `right' eigenfunction, defined by
\be
\cL_k r_k=\Lambda_k r_k.
\ee
We also denote by $l_k$ its `left' eigenfunction, defined by
\be
\cL_k^\dag l_k=\Lambda_k l_k,
\ee
where $\cL_k^\dag$ is the dual of $\cL_k$ with respect to the Lebesgue or counting measure. These eigenfunctions are defined, as usual, up to multiplicative constants, set here by imposing the following normalization conditions:
\be
\int_{\cE} l_k(x)dx=1\qquad\textrm{and}\qquad\int_{\cE} l_k(x)r_k(x)dx=1.
\label{eq:norm}
\ee
For the remaining, we also assume that the initial measure $\mu_0$ of $X_t$ is such that
\be
\int_\cE \mu_0(dx)\, r_k(x)<\infty,
\label{eq:condf}
\ee
and that there is a gap $\Delta_{k}$ between the first two largest eigenvalues resulting from the Perron--Frobenius Theorem. Under these assumptions, the semi-group generated by $\cL_k$ admits the asymptotic expansion
\be
e^{t\cL_k}(x,y)=e^{t\Lambda_k}\left[r_k(x)l_k(y)+O(e^{-t\Delta_k})\right]
\label{eq:asysg}
\ee
as $t\ra\infty$. Applying this result to the Feynman--Kac formula
\be
\Ex_{\mu_0}[e^{kTA_T}\delta (X_T-y)]=\int_{\cE} \mu_0(dx_0)\, e^{T\cL_k} (x_0,y),
\label{eq:dvl}
\ee
obtained by integrating (\ref{eq:ptilt}) with $C=\delta(X_{T}-y)$, yields
\be
\Ex_{\mu_0}[e^{kTA_T}\delta (X_T-y)]= e^{T\Lambda_k} \int_{\cE} \mu_0(dx_0) \left[ r_k(x_0) l_k(y) +O(e^{-t\Delta_k})\right].
\label{eq:t-T}
\ee
From this relation, we then deduce the following representations of the spectral elements $\Lambda_k$, $r_k$, and $l_k$; a further representation for the product $r_kl_k$ will be discussed in the next subsection.
\begin{itemize}
\item Dominant eigenvalue $\Lambda_{k}$:
\be
\Lambda_{k}=\lim_{T\ra\infty}\frac{1}{T}\ln \Ex_{\mu_0}[e^{kTA_T}]
\label{eq:fgc-1}
\ee
for all $\mu_0$ such that (\ref{eq:condf}) is satisfied.

\item Right eigenfunction $r_k$:
\be
r_k(x_0)=\lim_{T\ra\infty} \, e^{-T\Lambda_k} \Ex_{x_0}[e^{kTA_T}]
\label{eq:rk}
\ee
for all initial condition $x_0$.

\item Left eigenfunction $l_k$:
\be
l_k(y) =\lim_{T\ra\infty} \frac{\Ex_{\mu_0}[e^{kTA_T} \delta(X_T-y)]}{\Ex_{\mu_0}[e^{kTA_T}]}
\label{eq:rsl}
\ee
for all $\mu_0$ such that (\ref{eq:condf}) is satisfied.

\end{itemize}

With these results, we can already build a path measure from $d\Pr_{\cL_k,\mu_0,T}$, which is asymptotically equivalent to the canonical path measure. Indeed, it is clear from (\ref{eq:fgc-1}) that
\be
\lim_{T\ra\infty} \frac{1}{T} \ln \left(e^{-T\Lambda_k}\frac{d\Pr_{\cL_k,\mu_0,T}}{d\Pr_{k,\mu_0,T}^\cano}\right)=0
\ee
almost everywhere, so that
\be
d\Pr_{k,\mu_0,T}^\cano \asymp e^{-T\Lambda_k} d\Pr_{\cL_k,\mu_0,T}.
\label{eq:preeq}
\ee
We will see in the next section how to integrate the constant term $e^{-T\Lambda_k}$ into a Markovian measure, so as to obtain a Markov process which is conservative and equivalent to the canonical ensemble.
For now, we close this subsection with two remarks:
\begin{itemize}
\item The right-hand side of (\ref{eq:fgc-1}) is known in large deviation theory as the \emph{scaled cumulant generating function} (SCGF) of $A_T$. The rate function $I$ can be obtained from this function using the G\"artner-Ellis Theorem \cite{dembo1998,hollander2000,ellis1985}, which states (in its simplest form) that, if $\Lambda_k$ is differentiable, then $A_T$ satisfies the LDP with rate function $I$ given by the Legendre-Fenchel transform of $\Lambda_k$:
\be
I(a)=\sup_{k}\{ka-\Lambda_k\}.
\label{eq:ge}
\ee
For pure jump processes on a finite space, the differentiability of $\Lambda_k$ follows from the implicit function theorem and the fact that $\Lambda_k$ is a simple zero of the characteristic polynomial. For cases where $\Lambda_k$ is nondifferentiable, see Sec.~4.4 of \cite{touchette2009}.

\item The cloning simulation methods \cite{giardina2006,lecomte2007a,tailleur2009} mentioned in the introduction can be interpreted as algorithms that generate the non-conservative process $\cL_k$ and obtain the SCGF $\Lambda_k$ by estimating the rate of growth or decay of its (non-normalized) measure, identified as $\Ex_{\mu_0}[e^{kTA_T}]$. An alternative method for simulating large deviations is transition path sampling, which attempts to directly sample paths according to $\Pr_{k,\mu_0,T}^{\cano}$ \cite{dellago2003,dellago2006,dellago2009,vanden-eijnden2006b}.
\end{itemize}

\subsection{Marginal canonical density}

Equation (\ref{eq:rsl}) can be reformulated in terms of the canonical path measure as 
\be
l_k(y)= \lim_{T\ra\infty}\int d\Pr_{k,\mu_0,T}^{\cano}(\om)\, \delta(X_{T}(\om)-y).
\label{eq:intl}
\ee
This gives a physical interpretation of the left eigenfunction as the limit, when $T$ is large, of the marginal probability density of the canonical ensemble at the final time $t=T$. If we calculate this marginal for $t\in[0,T[$ and let $t\ra\infty$ after taking $T\ra\infty$, we obtain instead
\be
l_k(y)r_k(y)=\lim_{t\ra\infty}\lim_{T\ra\infty}\int d\Pr_{k,\mu_0,T}^{\cano}(\om)\, \delta(X_{t}(\om)-y).
\label{eq:intl-2}
\ee
The product $r_k l_k$ is thus the large-time marginal probability density of the canonical process taken over the infinite time interval. We will see in Sec.~\ref{sec:emp} that the same product corresponds to the invariant density of the driven process. 

To prove (\ref{eq:intl-2}), take $C=\delta(X_{t}-y)$ with $t<T$ in (\ref{eq:ptilt}) and integrate to obtain
\be
\Ex_{\mu_0}[e^{kTA_T}\delta(X_{t}-y)] = 
\int_{\cE} \mu_0(dx_0)\, e^{t\cL_k} (x_0,y)\; (e^{(T-t)\cL_k}1)(y).
\label{eq:FK2}
\ee
Now, take the limit $T\ra\infty$ to obtain
\be
\lim_{T\ra\infty} \, e^{-T\Lambda_k} \Ex_{\mu_0} [e^{kTA_T} \delta(X_t-y)] = 
\int_{\cE} \mu_0(dx_0)\, e^{t\cL_k}(x_0,y)\; e^{-t\Lambda_k}\, r_k(y),
\label{eq:limFK2}
\ee
which can be rewritten with (\ref{eq:t-T}) as
\begin{eqnarray}
\lim_{T\ra\infty} \frac{\Ex_{\mu_0}[e^{kTA_T}\delta(X_t-y)]}{\Ex_{\mu_0}[e^{kTA_T}]} 
= \frac{\displaystyle\int_{\cE} \mu_0(dx_0)\, e^{t\cL_k}(x_0,y)\; e^{-t\Lambda_k}\, r_k(y)}{\displaystyle\int_{\cE} \mu_0(dx_0)\, r_k(x_0)}
\label{eq:limFK22}
\end{eqnarray}
assuming (\ref{eq:condf}). Finally, take the limit $t\ra\infty$ to obtain
\be
\lim_{t\ra\infty} \lim_{T\ra\infty} \frac{\Ex_{\mu_0}[ e^{kTA_T}\delta(X_{t}-y)]}{\Ex_{\mu_0}[e^{kTA_T}]} 
= l_k(y)r_k(y),
\label{eq:intstolr}
\ee
which can be rewritten with the canonical measure as (\ref{eq:intl-2}). A similar proof applies to (\ref{eq:intl}); see Appendix B of \cite{lecomte2007b} for a related discussion of these results.

The result of (\ref{eq:intl-2}) can actually be generalized in the following way: instead of taking $t\in[0,T[$ and letting $t\ra\infty$ after $T\ra\infty$, we can scale $t$ with $T$ by choosing  $t=c(T)$ such that
\be
\lim_{T\ra\infty} c(T)=\infty\qquad \text{and}\qquad \lim_{T\ra\infty} T-c(T)=\infty.
\ee
In this case, it is easy to see from (\ref{eq:limFK2})-(\ref{eq:intstolr}) that we obtain the same result, namely,
\be
l_k(y)r_k(y)=\lim_{T\ra\infty}\int d\Pr_{k,\mu_0,T}^{\cano}(\om)\, \delta(X_{c(T)}(\om)-y).
\label{eq:intl-2-2}
\ee
In particular, we can take $c(T)=(1-\epsilon) T$ with $0<\epsilon<1$ to get $t$ as close as possible to $T$, without reaching $T$. This will be used later when considering the equivalence of the driven process with the canonical path measure.

Note that there is no contradiction between (\ref{eq:intl}) and (\ref{eq:intl-2}), since for $t\leq T$, 
\begin{eqnarray}
\int d\Pr_{k,\mu_0,T}^\cano(\om)\, \delta(X_t(\om)-y)
&=&\frac{\displaystyle\int_{\cE} \mu_0(dx_0)\, e^{t\cL_k}(x_0,y)\; (e^{(T-t)\cL_k}1)(y)}{\displaystyle\int_{\cE} \mu_0(dx_0)\, (e^{T\cL_k}1)(x_0)}\nonumber\\
&\neq& \frac{\displaystyle\int_{\cE} \mu_0(dx_0)\, e^{t\cL_k}(x_0,y)}{\displaystyle\int_{\cE} \mu_0(dx_0)\, (e^{t\cL_k}1)(x_0)}
=\int d\Pr_{k,\mu_0,t}^\cano (\om)\, \delta(X_{t}(\om)-y).
\label{eq:rem}
\end{eqnarray}
The fact that the left-most and right-most terms are not equal arises because the canonical measure is defined globally (via $A_T$) for the whole time interval $[0,T]$, so that the marginal of the canonical measure at time $t$ depends on times after $t$, as well as the end-time $T$. We will study in more detail the source of this property in Sec.~\ref{sec:emp} when proving that the canonical path measure is a non-homogeneous Markov process that explicitly depends on $t$ and $T$.

\section{Generalized Doob transform}
\label{sec:gdt}

We define in this section the generalized Doob transform that will be used in the next section to define the driven process. We also review the conditioning problem considered by Doob to understand whether the case of large deviation conditioning can be analyzed within Doob's approach. Two examples will be considered: first, the original problem of Doob involving the conditioning on leaving a domain via its boundary and, second, a `punctual' conditioning at a deterministic time. In each case, we will see that the generator of the process realizing the conditioning is a particular case of Doob's transform, but that the random variable underlying the conditioning is, in general, different  from the random variables $A_T$ defined before. 

\subsection{Definition}
\label{sub:htransform}

Let $h$ be a strictly positive function on $\cE$ and $f$ an arbitrary function on the same space. We call the \emph{generalized Doob transform} of the process $X_t$ with generator $L$ the new process with generator
\be
L^{h,f}\equiv h^{-1}Lh-f.
\label{eq:dg}
\ee
In this expression, $h^{-1}Lh$ must be understood as the composition of three operators: the multiplication operator by $h^{-1}$, the operator $L$ itself, and the multiplication operator by $h$. Moreover, the term $f$ represents the multiplication operator by $f$, so that the application of $L^{h,f}$ on some function $r$ yields
\be
(L^{h,f}r)(x)= h^{-1}(x)\, (Lhr)(x)-f(x)r(x).
\ee

We prove in Appendix~\ref{app:chdoob} that the generalized Doob transform of $L$ is indeed the generator of a Markov process, whose path measure $\Pr_{L^{h,f},\mu_0,T}$ is absolutely continuous with respect to the path measure $\Pr_{L,\mu_0,T}$ of $X_t$ and whose Radon--Nikodym derivative is explicitly given by 
\be
\frac{d\Pr_{L^{h,f},\mu_0,T}}{d\Pr_{L,\mu_0,T}}(\om)=h^{-1}(X_0)\exp\left(-\int_{0}^{T}f(X_{t})\, dt\right)h(X_{T}).
\label{eq:simi}
\ee
In the following, we will also use time-dependent functions $h_t$ and $f_t$ to transform $L$ \cite{chetrite2011}. In this case, the generalized Doob transform is a non-homogeneous process with path measure given by
\be
\frac{d\Pr_{L^{h,f},\mu_0,T}}{d\Pr_{L,\mu_0,T}}(\om)
=
h_0^{-1}(X_0) \exp\left(-\int_{0}^{T} (f_t+h_t^{-1}\partial_t h_t)(X_t)\, dt\right) h_T(X_T).
\label{eq:rnonh}
\ee

It is important to note that the transformed process with generator $L^{h,f}$ is Markovian, but not necessarily conservative, which means that its dominant eigenvalue is not necessarily zero. If we require conservation (zero dominant eigenvalue), it is sufficient that we choose $f=h^{-1}(Lh)$, in which case (\ref{eq:dg}) becomes
\be
L^h=h^{-1}Lh-h^{-1}(Lh),
\label{eq:sim}
\ee
while (\ref{eq:simi}) reduces to
\be
\frac{d\Pr_{L^{h},\mu_0,T}}{d\Pr_{L,\mu_0,T}}(\om)=h^{-1}(X_0)\exp\left(-\int_{0}^{T}h^{-1}(X_t)\; (Lh)(X_t)\, dt\right)h(X_{T}).
\label{eq:simi2}
\ee
Moreover, in the time-dependent case, (\ref{eq:rnonh}) becomes
\be
\frac{d\Pr_{L^{h},\mu_0,T}}{d\Pr_{L,\mu_0,T}}(\om)
=
h_0^{-1}(X_0) \exp\left(-\int_{0}^{T}dt\left(h_{t}^{-1}(Lh_t)+h_{t}^{-1}\partial_{t}h_{t}\right)(X_t)\right) h_T(X_T).
\label{eq:nonstat}
\ee

Specializing to specific processes, it is  easy to see that the generalized Doob transform of a pure jump process with transition rates $W(x,dy)$ is also a pure jump process with modified transition rates
\be
W^{h}(x,dy)=h^{-1}(x)W(x,dy)h(y)
\label{eq:htps}
\ee
for all $(x,y)\in\cE^2$. Similarly, it can be shown that the generator of the generalized Doob transform of a diffusion with generator $L$ is
\be
L^h=L+(\nabla\ln h) D\nabla,
\label{adddrift}
\ee
where the product involving $D$ is interpreted, as before, according to (\ref{eq:dp}). The generalized Doob transformed process is thus a diffusion with the same noise as the original diffusion, but with a modified drift
\be
F^h=F+D\,\nabla\ln h.
\label{eq:tdpe-2}
\ee
The proof of this result is given in Appendix~\ref{app:pcarc} and follows by re-expressing the generalized Doob transform of (\ref{eq:sim}) as
\begin{equation}
L^h=L+h^{-1}\Gamma(h,\cdot),
\label{eq:simicar}
\end{equation}
where $\Gamma$ is the so-called `squared field' operator,\footnote{From the French `op\'erateur carr\'e du champs'.} which is a symmetric bilinear operator defined for all $f$ and $g$ on $\cE$ as
\be
\Gamma(f,g)\equiv (Lfg)-f(Lg)-(Lf)g.
\label{eq:carrechamp}
\ee

Mathematical properties and applications of the generalized Doob transform have been studied by Kunita \cite{kunita1969}, It\^o and Watanabe \cite{ito1965}, Fleming and collaborators (see \cite{fleming2006} and references cited therein), and have been revisited recently by Palmowski and Rolski \cite{palmowski2002} and Diaconis and Miclo \cite{diaconis2009}. From the point of view of probability theory, the Radon--Nikodym derivative associated with this transform is an example of exponential martingale. The generalized Doob transform also has interesting applications in physics: it appears in the stochastic mechanics of Nelson \cite{meyer1985} and underlies, as shown in \cite{chetrite2011}, the classical fluctuation--dissipation relations of near-equilibrium systems \cite{bernard1959,callen1951,kubo1966,risken1996}, and recent generalizations of these relations obtained for nonequilibrium systems \cite{lippiello2008,speck2006,baiesi2009,baiesi2009b,baiesi2010,seifert2010,chetrite2008,chetrite2009}. The work of \cite{chetrite2011} shows moreover that the exponential martingale (\ref{eq:simi2}) verifies a non-perturbative general version of these relations, which also include the fluctuation relations of Jarzynski \cite{jarzynski1997} and Gallavotti-Cohen \cite{evans1993,gallavotti1995,gallavotti1995a}.

\subsection{Historical conditioning of Doob}
\label{sub:hdt}

The transform considered by Doob is a particular case of the generalized transform (\ref{eq:dg}), obtained for the constant function $f(x)\equiv\lambda$ and for a so-called \textit{$\lambda$-excessive} function $h$ verifying $Lh\leq\lambda h$. For these functions, the Doob transformed process is a non-conservative process of generator 
\be
L^{h,\lambda}=h^{-1}Lh-\lambda,
\label{eq:dh}
\ee
and path measure
\be
\frac{d\Pr_{L^{h,\lambda},\mu_0,T}}{d\Pr_{L,\mu_0,T}}(\om)=h^{-1}(X_0)e^{-T\lambda}h(X_T).
\label{eq:simivp-1}
\ee
When $(Lh)=\lambda $, $h$ is said to be $\lambda$-invariant. If we also have $\lambda=0$, then $h$ is called a \emph{harmonic} function \cite{doob1984,doob1957,rogers2000}, and the process described by $L^h=h^{-1}Lh$ is conservative with path measure
\be
\frac{d\Pr_{L^h,\mu_0,T}}{d\Pr_{L,\mu_0,T}}(\om)=h^{-1}(X_0)h(X_T).
\label{eq:doobharmo}
\ee
In the time-dependent case, the harmonic condition $Lh=0$ is replaced by
\be
(\partial_t+L_t)h_t=0,
\label{eq:sth}
\ee
which yields, following (\ref{eq:rnonh}) and (\ref{eq:nonstat}),
\be
\frac{d\Pr_{L^{h},\mu_0,T}}{d\Pr_{L,\mu_0,T}}(\om)=h_0^{-1}(X_0)h_T(X_T).
\label{eq:tdd}
\ee
In this case, $h_t$ is said to be \emph{space--time harmonic} \cite{doob1984,doob1957,rogers2000}.
Applications of these transforms have appeared since Doob's work in the context of various conditionings of Brownian motion, including the Gaussian and Schr\"odinger bridges mentioned in the introduction, in addition to non-colliding random walks related to Dyson's Brownian motion and random matrices \cite{dyson1962,grabiner1999,oconnell2003}.

The original problem considered by Doob, leading to $L^h$, is to condition a Markov process $X_t$ started at $X_0=x_0$ to exit a certain domain $D$ via a subset of its boundary $\partial D$. To be more precise, assume that the boundary of $D$ can be decomposed as $\partial D=B\cup C$ with $B\cap C=\emptyset$, and condition the process to exit $D$ via $B$. In this case, the path measure of the conditioned process can be written as 
\be
d\Pr_{x_0,T}\{\om|\cB\}=d\Pr_{L,x_0,T}(\om) \frac{\Pr_{L,x_0}\{\cB|\cF_T\}}{\Pr_{L,x_0}\{\cB\}},
\label{eq:mc}
\ee
where $\Pr_{L,x_0,T}$ is the path measure of the process started at $x_0$, $\cB=\{\tau_B\leq \tau_C\}$ is the conditioning event expressed in terms of the exit times,
\be
\tau_B\equiv \inf \{t: X_t\in B\},\qquad \tau_C\equiv \inf \{t:X_t\in C\},
\ee
and $\cF_T=\sigma\{X_t(\om):0\leq t\leq T\}$ is the natural filtration of the process up to the time $T$.  

The conditional path measure (\ref{eq:mc}) is similar to the microcanonical path measure (\ref{eq:micro}) and can be expressed in the form
\be
d\Pr_{x_0,T}\{\om|\cB\}=d\Pr_{L,x_0,T}(\om) M_{[0,T]},
\label{eq:mcp}
\ee
where
\be
M_{[0,T]}=\frac{\Pr_{L,x_0}\{\cB|\cF_T\}}{\Pr_{L,x_0}\{\cB\}}
\label{eq:mcp2}
\ee
to emphasize that it is a `reweighing' or `penalization' \cite{roynette2009} of the original measure of the process with the \emph{weighting function} $M_{[0,T]}$. To show that this reweighing gives rise to a Doob transform for the exit problem, let
\be
h(x)=P_{x}\{\tau_B\leq\tau_C\}.
\label{eq:h}
\ee
This function is harmonic, since $(Lh)=0$ by Dynkin's formula \cite{rogers2000}. Moreover, using the strong Markov property, we can use this function to express the weighting function (\ref{eq:mcp2}) as
\be
M_{[0,T]}=\frac{h(X_{\min(T,\tau_{B},\tau_{C})})}{h(X_0)}.
\label{eq:M1S-1}
\ee
For $T\leq \min(\tau_{B},\tau_{C})$, we therefore obtain
\be
d\Pr_{x_0,T}\{\om|\cB\}=h^{-1}(x_0)d\Pr_{L,x_0,T}(\om)h(X_{T}).
\label{eq:doobmicro-1-1}
\ee
which has the form of (\ref{eq:doobharmo}). The next example provides a simple application of this result. 

\begin{example}
Consider the Brownian or Wiener motion $W_t$ conditioned on exiting the set $A=\{ 0,\ell\} $ via $B=\{\ell\} $. The solution of $(Lh)=0$ with the boundary conditions $h(0)=0$ and $h(\ell)=1$ gives the harmonic function $h(x)=x/\ell$, which implies from (\ref{eq:tdpe-2}) that the drift of the conditioned process is $F^h(x)=1/x$. The conditioned process is thus the Bessel process:
\be
dX_{t}=\frac{1}{X_{t}}dt+dW_{t}.
\ee
Note that the drift of the conditioned process is independent of $\ell$, which means by taking $\ell\ra\infty$ that the Bessel process is also the Wiener process conditioned never to return at the origin. This is expected physically, as $F^h$ is a repulsive force at the origin which prevents the process from approaching this point.
\end{example}

As a variation of Doob's problem, consider the conditioning event
\be
\cB_T=\{ X_T\in B_T\},
\ee
where $B_{T}$ is a subset of $\cE$ that can depend on $T$. This event is a particular case of $A_T$ obtained with $f=0$ and $g=1$, so that $A_T=X_T/T$ assuming $X_0=0$.  Its associated weighting function takes the form
\be
M_{[0,T']} 
= \frac{\Pr_{L,x_0}\{\cB_T|\cF_{T'}\}}{\Pr_{L,x_0}\{\cB_T\}}
= \frac{\Pr_{L,x_0}\{\cB_T|X_{T'}\}}{\Pr_{L,x_0}\{\cB_T\}},
\label{eq:fp-1-2}
\ee
for $T'\leq T$. Defining the function
\be
h_{T'}(X_{T'})\equiv \Pr_{L,x_0}\{\cB_T|X_{T'}\}=\int_{B_T} P_{T-T'}(X_{T'},dy),
\label{eq:fch}
\ee
we then have
\be
M_{[0,T']}=\frac{h_{T'}(X_{T'})}{h_0(X_0)}.
\label{eq:M2S-1}
\ee
Moreover, from the backward Kolmogorov equation, we find that $h$ is space--time harmonic, as in (\ref{eq:sth}). Therefore, the path measure of $X_t$ conditioned on $\cB_T$ also takes the form of a Doob transform,
\be
d\Pr_{x_0,T'}\{\om|\cB_T\}  =  d\Pr_{L,x_0,T'}(\om)\, h_0^{-1}(x_0)h_{T'}(X_{T'}),
\label{eq:doob}
\ee
but now involves a time-dependent space--time harmonic function.\footnote{Note that the probability of $\cB_T$ can vanish as $T'\ra\infty$, for example, if $X_t$ is transient. In this case, (\ref{eq:fch}) vanishes as $T'\ra\infty$, so that (\ref{eq:doob}) becomes singular in this limit.} 

The next two examples apply this type of punctual conditioning to define bridge versions of the Wiener motion and the Ornstein--Uhlenbeck process.

\begin{example}[Brownian bridge]
Let $W_t|W_T=0$ be the Wiener motion $W_t$ conditioned on reaching $0$ at time $T$. The Kolmogorov equation, which in this case is simply the classical diffusion equation with $L=\Delta/2$, yields the Gaussian transition density of $W_t$ as the space--time harmonic function:
\be
h_{t}(x)=e^{(T-t)L}(x,0)=\frac{1}{\sqrt{2\pi(T-t)}} \exp\left(-\frac{x^2}{2(T-t)}\right),\qquad 0\leq t<T.
\ee
From (\ref{eq:tdpe-2}) and (\ref{eq:fch}), we then obtain, as expected, that $W_t|W_T=0$ is the Brownian bridge evolving according to
\be
dX_t=-\frac{X_t}{T-t}dt+dW_t
\label{eq:pb}
\ee
for $0\leq t<T.$ The limit $T\ra\infty$ recovers the Wiener process itself as the conditioned process.
\end{example}

\begin{example}[Ornstein--Uhlenbeck bridge]
\label{par:OU}
Consider now the Ornstein--Uhlenbeck process, 
\be
dX_t=-\gamma X_tdt+\sigma dW_t
\label{eq:OU}
\ee
with $\gamma>0$ and $\sigma>0$, conditioned on the event $X_T=Ta$. Using the propagator of this process, 
\be
P_t(x,y)=\sqrt{\frac{\gamma}{\pi\sigma^2(1-e^{-2\gamma t})}}\;
\exp\left(-\frac{\gamma}{\sigma^2}\frac{(y-e^{-\gamma t}x)^2}{1-e^{-2\gamma t}}\right),
\label{eq:PTOU-2}
\ee
we obtain from (\ref{eq:fch}),
\be
h_t(x)=e^{(T-t)L}(x,0)=\sqrt{\frac{\gamma}{\pi \sigma^2 (1-e^{-2\gamma (T-t)})}}\;
\exp\left(-\frac{\gamma}{\sigma^{2}} \frac{ (x-Ta\, e^{\gamma(T-t)})^2}{e^{2\gamma(T-t)}-1}\right).
\label{eq:hou}
\ee
With (\ref{eq:tdpe-2}), we then conclude that $X_t|X_T=aT$ is the non-homogeneous diffusion
\be
dX_t=-\gamma X_tdt+F_{T}(X_t,t)dt+\sigma dW_{t},\qquad 0\leq t<T,
\label{eq:pgou}
\ee
with added time-dependent drift
\be
F_{T}(x,t)=-2\gamma\frac{x-Tae^{\gamma(T-t)}}{e^{2\gamma(T-t)}-1}.
\label{eq:addedd}
\ee

The relation between this drift and the conditioning is interesting. Since
\be
\lim_{t\ra T} F_T(x,t)=
\left\{
\begin{array}{lll}
\infty & & x<aT \\
\gamma a T& & x=aT\\
-\infty & & x>aT,
\end{array}
\right.
\ee
points away from the target $x=aT$ are infinitely attracted toward this point as $t\ra T$, which leads $X_t$ to reach $X_T=aT$. This attraction, however, is all concentrated near the final time $T$, as shown in Fig.~\ref{fig1}, so that the conditioning $X_T=aT$ affects the Ornstein--Uhlenbeck process mostly at the boundary of the time interval $[0,T]$ and marginally in the interior of this interval. Taking the limit $T\ra\infty$ pushes the whole effect of the conditioning to infinity, so that care must be taken when interpreting this limit. It is clear here that we cannot conclude that, because $F_\infty(x,t)=0$ for $t< \infty$, the conditioned process is the Ornstein--Uhlenbeck process itself.

\begin{figure}
\includegraphics{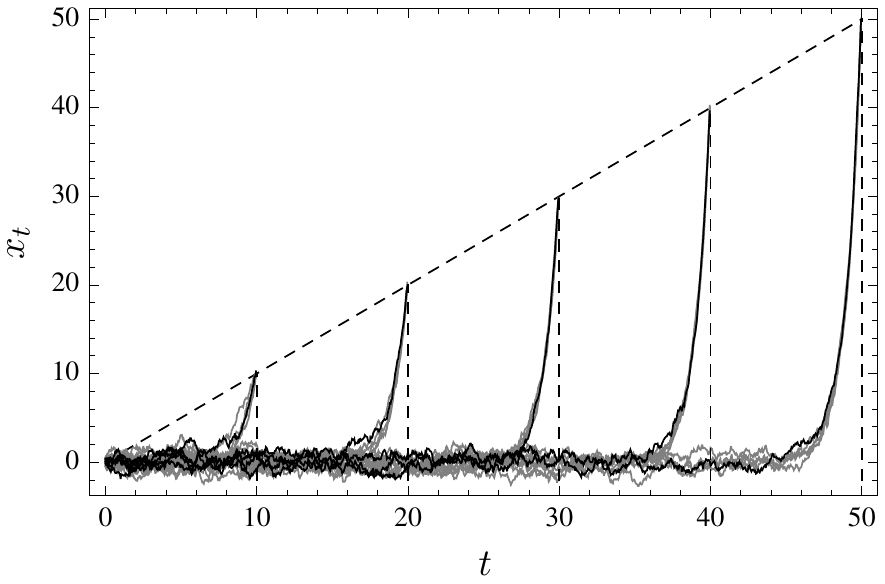}
\label{fig1}
\caption{Sample paths $\{x_t\}_{t=0}^T$ of the Ornstein--Uhlenbeck process conditioned on the final point $X_T=aT$ for $T\in \{10,20,30,40,50\}$. Parameters: $\gamma=1$, $\sigma = 1$, $a=1$. Black curves highlight one of five sample paths generated for each $T$. The conditioning mostly affects, as clearly seen, the dynamics  only near the final time $T$, over a constant time-scale, inferred from (\ref{eq:addedd}), to be roughly given by $1/\gamma$.}
\end{figure}

This boundary behavior of the conditioning will be discussed later. Interestingly, this behavior does not arise for the Wiener motion, obtained with $\gamma=0$ and $\sigma=1$. In this case, the conditioned process is
\be
dX_t=-\frac{X_t-Ta}{T-t}dt+dW_{t}
\label{eq:pcw}
\ee
and converges to 
\be
dX_t=adt+dW_t
\label{eq:pew}
\ee
in the limit $T\ra\infty$. Thus, the conditioning $X_T=aT$ is effected by an added drift $a$, which affects the dynamics of the process over the complete interval $[0,T]$.
\end{example}

We return at this point to our original problem of representing in terms of a conservative Markov process the microcanonical path measure $\Pr_{a,\mu_0,T}^\micro$ associated with the large deviation conditioning $X_t|A_T=a$. Following the preceding examples, the obvious question arises as to whether this measure can be obtained from a `normal' Doob transform involving a suitably chosen function $h$. The answer is, no, for essentially two reasons:
\begin{itemize}
\item Since $A_T$ depends on the whole time interval $[0,T]$ and not, as in the examples above, on a `punctual' random time $\tau\leq T$ or a deterministic time $T$, the weighting function associated with the large deviation conditioning (\ref{eq:microb}) cannot be expressed as in (\ref{eq:M1S-1}) or (\ref{eq:M2S-1}). What must be considered for this type of conditioning is an \emph{approximate} and \emph{asymptotic} form of equivalence, which essentially neglects the boundary terms $h(X_0)$ and $h(X_T)$, as well as sub-exponential terms in $T$.

\item There does not seem to be a way to prove the equivalence of the microcanonical path measure with a Markov measure starting directly from the definitions of the former measure, the associated weighing function, and the conditioning observable $A_T$. Here, we prove this equivalence indirectly via the use of the canonical path measure.
\end{itemize}
These points are discussed in more detail in the next section. 

\section{Driven Markov process}
\label{sec:emp}

We now come to the main point of this paper, which is to define a Markov process via the generalized Doob transform and prove its asymptotic equivalence with the conditioned process $X_t|A_T=a$. This equivalence is obtained, as just mentioned, by first proving the asymptotic equivalence of the path measure of the driven process with the canonical path measure, and by then proving the equivalence of the latter measure with the microcanonical path measure using known results about ensemble equivalence. Following these results, we discuss interesting properties of the driven process related to its reversibility and constraints satisfied by its transition rates (in the case of jump processes) or drift (in the case of diffusions). Some of these properties were announced in \cite{chetrite2013}; here, we provide their full proofs in addition to deriving new results concerning the reversibility of the driven process. Our main contribution is to treat the equivalence of the canonical and microcanonical path ensembles explicitly and derive conditions for this equivalence to hold. In previous works, the conditioned process is assumed to be equivalent to the driven process and, in some cases, wrongly interpreted as the canonical path ensemble.

\subsection{Definition}

We define the \emph{driven process} $Y_t$ by applying the generalized Doob transform to the generator $\cL_k$ of the non-conservative process considered in Sec.~\ref{sec:tiltedproc}, using for $h$ the right eigenfunction $r_k$, which is strictly positive on $\cE$ by Perron--Frobenius. We denote the resulting generator of $Y_t$ by $L_k$, so that in the notation of the generalized Doob transform (\ref{eq:sim}), we have
\be
L_k\equiv \cL_k^{r_k}=r_k^{-1}\cL_{k}r_k-r_k^{-1}(\cL_{k}r_k).
\label{eq:resultat}
\ee
Although the tilted generator $\cL_k$ is not conservative, $L_k$ is since $(L_k1)=0$. Moreover, we infer from (\ref{eq:simi}) that the path measure of this new process is related to the path measure of the non-conservative process by
\be
\frac{d\Pr_{L_k,\mu_0,T}}{d\Pr_{\cL_k,\mu_0,T}}=r_k^{-1}(X_0)\, e^{-T\Lambda_k}\, r_k(X_T),
\label{eq:simivppt}
\ee
which means, using (\ref{eq:ptilt'}), that it is related to the path measure of the original (conservative) process by
\be
\frac{d\Pr_{L_k,\mu_0,T}}{d\Pr_{L,\mu_0,T}}=\frac{d\Pr_{L_k,\mu_0,T}}{d\Pr_{\cL_k,\mu_0,T}} \frac{d\Pr_{\cL_k,\mu_0,T}}{d\Pr_{L,\mu_0,T}} =r_k^{-1}(X_0)\, e^{-T\Lambda_k}\, e^{kTA_T}\, r_k(X_T).
\label{eq:rndsim}
\ee

The existence and form of $L_k$ is the main result of this paper. Following the expressions of the tilted generator (\ref{eq:gtps}) and (\ref{eq:gtpd}), $L_k$ can also be re-expressed as
\be
L_k=r_k^{-1}\left.\cL_{k}\right|_{f=0}r_k+kf-\Lambda_k
\label{eq:alt}
\ee
to make the dependence on $f$ more explicit. We deduce from (\ref{eq:gtps}) and this result that the driven process associated with a pure jump process remains a pure jump process described by the modified rates
\be
W_{k}(x,dy)=r_k^{-1}(x)\, W(x,dy)\, e^{kg(x,y)}\, r_k(y)
\label{eq:pesp}
\ee
for all $(x,y)\in\cE^2$. For a pure diffusion $X_t$ described by the SDE (\ref{eq:diff}), the driven process $Y_t$ is a diffusion with the same noise as $X_t$, but with the following modified drift:
\be
F_k=F+D(kg+\nabla\ln r_k).
\label{eq:tdpe}
\ee
The proof of this result follows by explicitly calculating
\be
h^{-1}\cL_kh = \Fhat \cdot (\nabla+kg+\nabla\ln h)+\frac{1}{2}(\nabla+kg+\nabla\ln h)D(\nabla+kg+\nabla\ln h)+kf
\ee
for $h>0$ on $\cE$, so as to obtain
\be
\cL_k^h=\Fhat\cdot \nabla+\frac{1}{2}\nabla D \nabla+(kg+\nabla\ln h )D\nabla=L+(kg+\nabla\ln h )D\nabla.
\label{eq:simidiff}
\ee
Applying this formula to $h=r_k>0$, we obtain from (\ref{eq:gpd}) that $L_k$ is the generator of a diffusion with the same diffusion fields $\sigma_\alpha$ as $X_t$, but with the modified drift given in (\ref{eq:tdpe}). Note that this result carries an implicit dependence (via $r_k$) on the two functions $f$ and $g$ defining the observable $A_T$, in addition to the explicit dependence on $g$. 

\subsection{Equivalence with the canonical path ensemble}

The relations (\ref{eq:cano}), (\ref{eq:ptilt'}) and (\ref{eq:rndsim}) lead together to
\be
\frac{d\Pr_{L_k,\mu_0,T}}{d\Pr_{k,\mu_0,T}^\cano} 
 = 
\frac{d\Pr_{L_k,\mu_0,T}}{d\Pr_{L,\mu_0,T}}\, \frac{d\Pr_{L,\mu_0,T}}{d\Pr_{k,\mu_0,T}^\cano}
=
r_k^{-1}(X_0)r_k(X_T)\, e^{-T\Lambda_k}\, \Ex_{\mu_0}[e^{kTA_T}].
\label{eq:ratio3-1}
\ee
From the limit (\ref{eq:fgc-1}) associating the SCGF with $\Lambda_k$, we therefore obtain
\be
\lim_{T\ra\infty}\frac{1}{T}\ln \frac{d\Pr_{L_k,\mu_0,T}}{d\Pr_{k,\mu_0,T}^\cano} (\om)=0
\ee
for all paths, which shows that the path measure of the driven process is asymptotically equivalent to the canonical path measure. This means, as explained before, that the two path measures are logarithmically equivalent,
\be
d\Pr_{\cL_k^{r_k},\mu_0,T}\asymp  d\Pr_{k,\mu_0,T}^\cano,
\ee
so that, although they are not equal, their differences are sub-exponential in $T$ for almost all paths.

From this result, it is possible to show, with additional conditions, that the typical values of observables satisfying LDPs with respect to these measures are the same.\footnote{H. Touchette, in preparation, 2014.} However, because of the specific form of the canonical path ensemble, we can actually prove a stronger form of equivalence between this path measure and that of the driven process, which implies not only that observables have the same typical values, but also the same large deviations. 

This strong form of equivalence follows by noting that the canonical path ensemble represents a time-dependent Markov process. This is an important result, which does not seem to have been noticed before. The meaning of this is that, despite the global normalization factor $\Ex_{\mu_0}[e^{kTA_T}]$, the canonical measure defined in (\ref{eq:cano}) is the path measure of a non-homogeneous Markov process characterized by a time-dependent generator, denoted by $L^\cano_{k,t,T}$.\footnote{Time-dependent generators arise when considering probability kernels $P_s^t$ that depend on the times $s$ and $t$ between two transitions, and not just the time difference $t-s$, as considered in (\ref{CK}).} The derivation of this generator is presented in Appendix~\ref{app:gcpm}; the result is
\be
L_{k,t,T}^\cano\equiv\cL_k^{h_{t,T}}=h_{t,T}^{-1}\, \cL_k\, h_{t,T}-h_{t,T}^{-1}(\cL_k h_{t,T})
\ee
for all $t\in [0,T]$, where
\be
h_{t,T}(x)=(e^{(T-t)\cL_k}1)(x)
\ee
is space--time harmonic with respect to $\cL_k$ (see Appendix~\ref{app:gcpm}). Thus we see that the canonical measure is the generalized Doob transform of $\cL_k$ obtained, interestingly, with a time-dependent function $h_{t,T}$ involving $\cL_k$ itself. At the level of path measures, we then have
\be
d\Pr_{k,\mu_0,T}^\cano=d\Pr_{L_{k,\cdot,T}^\cano,\mu_0,T},
\label{eq:eqcand}
\ee
a result which should be understood in the sense of (\ref{eq:cyl}), with $L$ replaced by the time-dependent generator $L_{k,t,T}^\cano$ and the normal exponential replaced by a time-ordered exponential \cite{chetrite2011}.

To relate this result to the driven process, note that $(e^{(T-t)\cL_k}1)$ becomes proportional to $r_k$ as $T\ra\infty$, so that 
\be
\lim_{T\ra\infty} L_{k,t,T}^\cano=(\cL_k)^{r_k}\equiv L_k.
\label{eq:bg1}
\ee
Thus, although the process described by $d\Pr_{k,\mu_0,T}^\cano$ over the time interval $[0,T]$ is non-homogeneous for $T<\infty$, it becomes homogeneous inside this time interval as the final time $T$ diverges. Moreover, it converges in this limit to the driven process itself, which is by definition a homogeneous process. This holds for all $t\in [0,T[$ in the limit $T\ra\infty$; for the final time $t=T$, we obtain instead
\be
\lim_{T\ra\infty} L_{k,T,T}^\cano = \cL_k-(\cL_k1).
\label{eq:bg2}
\ee
Consequently, the convergence of the canonical process toward the driven process applies only in $[0,T[$; at the boundary of this time interval, the canonical process converges to a different homogeneous process with generator (\ref{eq:bg2}). This explains from the point of view of generators why we obtain two different limits for the marginal canonical density at $t<T$ and $t=T$, as seen in Sec.~\ref{sec:tiltedproc}.

This difference between the `interior' (or `bulk') and `boundary' regimes of a process is an important feature of our theory. In a sense, this theory can only characterize the `interior' of a process (exponentially tilted or conditioned), since we push the boundary to infinity, so to speak, and consider large deviation events that arise entirely from the `interior' regime. Given that the canonical and driven processes are the same in this `interior' regime, the large deviations of $A_T$  or any other observable satisfying an LDP must therefore also be the same for both processes.

To be more precise, consider an observable $B_T$ and assume that this observable satisfies an LDP with respect to the canonical path measure with rate function
\be
I_k(b)\equiv \lim_{T\ra\infty} -\frac{1}{T}\ln \Pr_{k,\mu_0,T}^\cano\{B_T\in [b,b+db]\}.
\label{eq:ldpcan1}
\ee
Let us write this LDP as
\be
I_k(b)= \lim_{T\ra\infty} \lim_{\epsilon\ra 0^+} -\frac{1}{T}\ln \Pr_{k,\mu_0,T}^\cano\{B_{(1-\epsilon)T}\in [b,b+db]\}.
\ee
If we assume that the fluctuations of $B_T$ arise from the combined effect of canonical fluctuations of $X_t$ over the whole interval $[0,T]$ and not just the end interval $[(1-\epsilon)T,T]$, we can invert the limits on $T$ and $\epsilon$ to obtain
\begin{eqnarray}
I_k(b) &=&\lim_{\epsilon\ra 0^+} \lim_{T\ra\infty} -\frac{1}{T}\ln \Pr_{k,\mu_0,T}^\cano\{B_{(1-\epsilon)T}\in db\}\nonumber\\
&=&\lim_{\epsilon\ra 0^+} \lim_{T\ra\infty} -\frac{1}{T}\ln \left.\Pr_{k,\mu_0,T}^\cano\right|_{[0,(1-\epsilon)T]}\{B_{(1-\epsilon)T}\in db\},
\label{eq:ldpcan2}
\end{eqnarray}
where $\left.\Pr_{k,\mu_0,T}^\cano\right|_{[0,(1-\epsilon)T]}$ represents the projection of $\Pr_{k,\mu_0,T}^\cano$ on $[0,(1-\epsilon)T]$, which is different in general from $\Pr_{k,\mu_0,(1-\epsilon)T}^\cano$. We know from our discussion above that this projection converges in the limit $T\ra\infty$ to the path measure of the driven process. If we further assume that this convergence carries over to $B_{(1-\epsilon)T}$, we can then write
\begin{eqnarray}
I_k(b) &=&\lim_{\epsilon\ra 0^+} \lim_{T\ra\infty} -\frac{1}{T}\ln \Pr_{L_k,\mu_0}\{B_{(1-\epsilon)T}\in db\}\nonumber\\
&=&\lim_{\epsilon\ra 0^+} (1-\epsilon)\lim_{T\ra\infty} -\frac{1}{(1-\epsilon)T}\ln \Pr_{L_k,\mu_0}\{B_{(1-\epsilon)T}\in db\}\nonumber\\
&=& \lim_{T\ra\infty} -\frac{1}{T}\ln \Pr_{L_k,\mu_0}\{B_{T}\in db\}.
\label{eq:ldpcan3}
\end{eqnarray}
Consequently, the LDP for $B_T$ in the canonical path ensemble implies an LDP for this random variable with respect the driven process with the \emph{same} rate function. 

This reasoning is valid, as stressed above, if the large deviations of $B_T$ and $B_{(1-\epsilon)T}$ are the same in the canonical path ensemble, that is, if these large deviations arise from the `interior' part of the measure and not from the boundary interval $[(1-\epsilon)T,T]$. In most cases of interest, this is verified, although there are pathological cases for which the large deviations actually arise at the boundary. The asymptotic limit of the Ornstein--Uhlenbeck process with $X_T=aT$, discussed in Sec.~\ref{sec:gdt}, is such a case, which we will come back to in Sec.~\ref{sec:apps}.

\subsection{Equivalence with the microcanonical path ensemble}

We now come back to the problem of characterizing $X_t|A_T=a$ as a Markov process by showing that the canonical and microcanonical path measures are asymptotically equivalent. This second level of equivalence is weaker than the previous one, for the simple reason that the microcanonical and canonical path measures have different supports. Moreover, the fact that $A_T$ does not fluctuate in the microcanonical path ensemble (by definition of the conditioning) but does, generally, in the canonical path ensemble shows that the large deviation properties of observables cannot be the same in general in both ensembles. However -- and this is the crucial observation for the problem of conditioning -- they can have the same typical values of observables, under conditions related to the convexity of the rate function $I(a)$ \cite{touchette2011b,touchette2014}. Moreover, the same conditions imply that the microcanonical and canonical path measures are asymptotically equivalent in the logarithmic sense. We discuss these levels of equivalence next, beginning with the one based on typical values.

As before, we assume that the conditioning observable $A_T$ satisfies the LDP with respect to the path measure $\Pr_{L,\mu_0,T}$ of the reference process $X_t$ with rate function $I(a)$. We then consider an observable $B_T$ and assume that it satisfies an LDP with respect to the microcanonical path measure $\Pr_{a,\mu_0,T}^\micro$ with rate function $J^a$, as well as an LDP with respect to the canonical path measure $\Pr_{k,\mu_0,T}^\cano$ with rate function $J_k$. We denote the set of global minima of $J^a$ by $\cB^a$ and the global minima of $J_k$ by $\cB_k$. Since rate functions vanish at their global minimizers \cite{ellis1985,touchette2009}, we can also write
\be
\cB^a=\{b:J^a(b)=0\},\qquad \cB_k=\{b:J_k(b)=0\}.
\ee
These zeros are called \emph{concentration points} in large deviation theory \cite{touchette2009}, since they correspond to the values of $B_T$ at which the microcanonical or canonical measure does not decay exponentially with $T$. If these sets are singleton sets, then their unique element correspond to the typical value of $B_T$ in the sense of the ergodic theorem \cite{ellis1985,touchette2009}. For example, if $\cB^a=\{b^*\}$ for a given value $a$ of $A_T$, then $B_T\ra b^*$ as $T\ra\infty$ with probability 1 with respect to the microcanonical path measure $\Pr_{a,\mu_0,T}^\micro$. A similar result can obviously be stated for the canonical ensemble. 

The equivalence problem in this context is to determine pairs $(a,k)$ for which $\cB^a=\cB_k$. Such pairs turn out to be determined by the convexity properties of $I(a)$. Denote by $\partial I(a)$ the subdifferential of $I$ at $a$. Except possibly at boundary points, $I$ is convex at $a$ if $\partial I(a)\neq\emptyset$, and is conversely nonconvex at $a$ if $\partial I(a)=\emptyset$ \cite{rockafellar1970}. With these notations, we have \cite{touchette2011b,touchette2014}:
\begin{itemize}
\item If $I$ convex at $a$, then $\cB^a=\cB_k$ for all $k\in\partial I(a)$. 

\item If $I$ is nonconvex at $a$, then $\cB^a\cap\cB_k=\emptyset$ for all $k\in\reals$. Thus, in this case, there is no $k\in\reals$ such that $\cB^a=\cB_k$.
\end{itemize}
The proof of these results, found in \cite{touchette2014}, relies on the following general relationship between the rate functions $J_k$ and $J^a$, which derives from the definitions of the microcanonical and canonical ensembles:
\be
J_k(b)=\inf_{a}\{J^a(b)+I(a)+\Lambda_k-ka\}.
\label{eq:ldtrep}
\ee
The idea of the proof is to relate the zeros of the two sides of (\ref{eq:ldtrep}), which define $\cB_k$ and $\cB^a$, by noting that $I(a)\geq ka-\Lambda_k$ with equality if and only if $I(a)$ is convex; see \cite{touchette2011b,touchette2014} for details.

A remarkable property of the microcanonical and canonical measures is that the convexity of $I(a)$ not only determines the equality of $\cB^a$ and $\cB_k$ for general observables, but also the logarithmic equivalence of these measures. This brings us to the second level of equivalence, expressed by the following results:
\begin{itemize}
\item If $I$ is convex at $a$, then for all $k\in\partial I(a)$,
\be
\lim_{T\ra\infty}\frac{1}{T}\ln \frac{d\Pr_{a,\mu_0,T}^\micro}{d\Pr_{k,\mu_0,T}^\cano} (\om)=0,
\ee
almost everywhere with respect to $\Pr_{a,\mu_0,T}^\micro$ and $\Pr_{k,\mu_0,T}^\cano$.

\item If $I$ is nonconvex at $a$, then there is no $k\in\reals$ for which the limit above vanishes.
\end{itemize}
The proof of these results also follows from the definitions of the microcanonical and canonical measures; see \cite{touchette2014}.

Our problem of large deviation conditioning can now be solved by linking all the results obtained. To recapitulate: 
\begin{enumerate}
\item \textbf{Driven--canonical measure equivalence:} Assuming the existence of $\Lambda_k$, $l_k$, and $r_k$, that the conditions (\ref{eq:norm}) and (\ref{eq:condf}) are satisfied, and that the spectrum of $\cL_k$ has a gap, we have that the driven process obtained from the generalized Doob transform (\ref{eq:resultat}) is such that
\be
d\Pr_{L_k,\mu,T}\asymp d\Pr_{k,\mu_0,T}^\cano.
\label{eq:conc1}
\ee

\item \textbf{Driven--canonical observable equivalence:} Any observable $B_T$ satisfying an LDP with respect to the canonical path measure also satisfies an LDP with respect to the law of the driven process with the \emph{same} rate function, provided that these LDPs are not related to boundary effects. In this case, the large deviations -- and by consequence the concentration points -- of $B_T$ are the same for both the canonical and driven processes.

\item \textbf{Canonical--microcanonical measure equivalence:} If $I(a)$ is convex, then
\be
d\Pr_{k,\mu_0,T}^\cano(\om) \asymp d\Pr_{a,\mu_0,T}^\micro(\om)
\label{eq:conc2}
\ee
for all $k\in\partial I(a)$, almost everywhere with respect to both measures.

\item \textbf{Canonical--microcanonical observable equivalence:} $B_T$ has in general different rate functions in the canonical and microcanonical path ensembles; however, its concentration points are the same in both ensembles when $I(a)$ is convex.\footnote{Equilibrium systems also have, in general, different fluctuations in the microcanonical and canonical ensembles, but have the same equilibrium states when they are equivalent.}
\end{enumerate}

We reach two conclusions from these results. The first, obtained by combining (\ref{eq:conc1}) and (\ref{eq:conc2}), is that if $I$ is convex at the conditioning value $a$, then
\be
d\Pr_{a,\mu_0,T}^\micro(\om) \asymp d\Pr_{L_k,\mu,T}(\om)
\ee
almost everywhere with respect to both measures for all $k\in\partial I(a)$. At the level of processes, we therefore write
\be
X_t|A_T=a\,\procequiv\, Y_t,
\ee
where $Y_t$ is the driven process with generator $L_k$ such that $k\in\partial I(a)$. The second conclusion, obtained from the points 2 and 4 above, is that $X_t|A_T=a$ and $Y_t$ have the same typical values of observables, provided that these observables concentrate in a large deviation sense in the long-time limit and that $k\in\partial I(a)$. It is in this sense that we say that the conditioned process $X_t|A_T=a$ is \emph{realized} or \emph{represented} by the driven process $Y_t$: the two processes may (and will in general) have different fluctuation properties, but they have the same typical or concentration properties in the stationary limit when $I(a)$ is convex. In a more physical but looser sense, we can picture them as describing the same long-time stochastic dynamics.

The next subsections discuss further properties of the driven process playing an important role for describing nonequilibrium systems. We list next several remarks that relate more specifically to its equivalence with the conditioned process:
\begin{itemize}

\item If $A_T$ has a unique concentration point $a^*$, then it should be expected that
\be
X_t|A_T=a^*\, \procequiv \, X_t,
\ee
since $A_T\ra a^*$ in the limit $T\ra\infty$, so that this value is `naturally' realized by $X_t$. This follows from our results by noting that $0\in \partial I(a^*)$, $\Lambda_0=0$ and $r_0=1$ up to a constant, so that $L_{k=0}=L$ in general, and $F_{k=0}=F$ for diffusions. Hence, conditioning on a typical value of the process does not modify it in the asymptotic limit.

\item The conditioning $A_T=a$ is realized by the driven process as a typical value of $A_T$ in the stationary limit. That is, $A_T\ra a$ as $T\ra\infty$ with probability 1 with respect to the law of $Y_t$. This follows simply by taking $B_T=A_T$.

\item The equivalence of $\cB^a$ and $\cB_k$ also implies, in the case where these sets are singleton sets, that bounded functions $C(B_T)$ have the same expectation in the driven and conditioned processes as $T\ra\infty$. In other words, equivalence of concentration points also implies, in the case of unique concentration points, equality of expectations.

\item If $I(a)$ is convex and differentiable, then $\partial I(a)=\{I'(a)\}$, so that the value $k$ achieving equivalence is given by $k=I'(a)$. In the case where $I(a)$ is strictly convex, we also have by Legendre duality that $k$ is such that $\Lambda_k'=a$ \cite{touchette2009}. These results are large deviation analogs of the thermodynamic relations connecting, respectively, the temperature with the derivative of the entropy and the energy with the derivative of the free energy \cite{touchette2009}.

\item Since equivalence is for all $k\in\partial I(a)$, there is possibly more than one driven process realizing the typical states of a conditioned process. This interesting result should arise whenever $I(a)$ has exposed (convex) corners at which $\partial I(a)$ is not a singleton; see \cite{gingrich2014} for an example.

\item Conversely to the above remark, there can be conditionings $X_t|A_T=a$ that admit no driven process if $I$ is nonconvex at $a$. We conjecture that such a case of nonequivalent processes arises whenever $X_t$ is not ergodic and switch between `phases' that cannot be represented by a single, homogeneous Markov process.

\item The driven process is a priori not unique: since boundary terms in path measures are negligible at the level of the logarithmic equivalence, one could apply an extra generalized Doob transform to $L_k$ by choosing a function $h>0$ such that $(L_k h)=0$. From the definition (\ref{eq:resultat}) of $L_k$, this is equivalent to $(\cL_k r_k h)=\Lambda_k r_k h$. Since $\Lambda_k$ is non-degenerate, $h$ must therefore be a multiplicative constant having no effect on the driven process.

\item In the case of diffusions with constant noise power $\sigma(x)=\sigma$, the low-noise limit $\sigma\ra 0$ yields for the driven process a deterministic differential equation for the (unique) fluctuation path characterizing the conditioning $A_T=a$. This can be used to recover known results from the Freidlin--Wentzell theory of fluctuation paths and instantons for noise-perturbed SDEs \cite{freidlin1984}.

\item The conditions leading to the equivalence of $X_t|A_T=a$ and $Y_t$ prevent many processes from being treated within our theory. Examples include L\'evy processes for which there are in general no LDPs ($I(a)=0$ everywhere or $\Lambda_k=\infty$), processes for which the LDP for $A_T$ may have a scaling or `speed' in $T$ different from $T$, as illustrated in the next section, in addition to processes with `condensation' transitions for which either $\Lambda_k\ra\infty$, $\cL_k$ is gapless or the condition (\ref{eq:condf}) is not satisfied; see \cite{merhav2010,jack2010b,harris2013b,szavits2014} for examples.
\end{itemize}

\subsection{Invariant density}

The driven process has an invariant density on $\cE$ corresponding to
\be
\rho_k(x)=l_k(x)r_k(x),
\label{eq:dipe}
\ee
which is normalized following (\ref{eq:norm}). This is proved directly from the definition (\ref{eq:resultat}) of the generator of the driven process, whose dual is
\be
L_k^{\dag}=r_k\cL_k^{\dag}r_k^{-1}-r_k^{-1}(\cL_kr_k),
\ee
so that 
\be
(L_k^{\dag}l_kr_k)=r_k(\cL_{k}^{\dag}l_k)-l_k(\cL_{k}r_k)=\Lambda_kl_kr_k-\Lambda_kl_kr_k=0.
\label{eq:pr}
\ee
If the driven process is ergodic, this invariant density is also the (unique) stationary density, in the sense that
\begin{equation}
\rho_k(y)=\lim_{t\ra\infty}\lim_{T\ra\infty} \int_\cE d\Pr_{L_k,\mu_0,T}(\om)\, \delta (Y_t(\om)-y)
\end{equation}
and
\be
\rho_k(y)= \lim_{T\ra\infty} \int_\cE d\Pr_{L_k,\mu_0,T}(\om)\, \delta (Y_T(\om)-y).
\label{eq:intl-2-1}
\ee
In this case, the stationary density is therefore the same independently of the time interval $[0,T]$ considered, contrary to the canonical ensemble measure which gives two different results for the two limit above; see again (\ref{eq:intl}) and (\ref{eq:intl-2}). 

Note that if the original process $X_t$ is ergodic with invariant density $\invrho$, then $\rho_{k=0}=\invrho$ because $l_0=\invrho$ and $r_k=1$. Moreover, if $\cL_k$ is self-dual (hermitian), then $l_k=r_k\equiv \psi_k$ so that $\rho_k=\psi_k^2$, in a clear analogy with quantum mechanics.

\subsection{Reversibility properties}

It is interesting physically to describe the class of conditioning observables $A_T$ for which the driven process is either reversible (equilibrium) or non-reversible (nonequilibrium). We study this problem here by deriving a functional equation involving $f$ and $g$, whose solution provides a necessary and sufficient condition for $\rho_k$ to be a reversible stationary density. This equation is hard to solve in general; a simpler form is obtained by assuming that the reference Markov process $X_t$ is reversible, which leads us to study the following question: Under what conditioning is the driven process $Y_t$ reversible given that $X_t$ is reversible? 

To answer this question, we first consider pure jump processes. For all $(x,y)\in\cE^{2}$, the relation (\ref{eq:pesp}) for the driven transition rates implies
\be
\frac{W_k(x,y)}{W_k(y,x)}=\left(\frac{r_k(y)}{r_k(x)}\right)^2 \frac{W(x,y)}{W(y,x)}\, e^{k[g(x,y)-g(y,x)]}.
\label{eq:ratio1}
\ee
Therefore, the driven process is reversible with respect to its invariant density $\rho_k$ if and only if the ratio above can be written as $\rho_k(y)/\rho_k(x)$, which yields, with the expression of $\rho_k$ shown in (\ref{eq:dipe}), a non-trivial functional equation for $f$ and $g$.

We can simplify this equation by assuming that the reference process is reversible, as in (\ref{eq:BDS}). The ratio (\ref{eq:ratio1}) then becomes 
\be
\frac{W_k(x,y)}{W_k(y,x)}=\left(\frac{r_k(y)}{r_k(x)}\right)^2\frac{\invrho(y)}{\invrho(x)}\, e^{k[g(x,y)-g(y,x)]}.
\ee
For the driven process to remain reversible, it is thus sufficient that there exists a `potential' function $h$ on $\cE$ such that 
\be
g(x,y)-g(y,x)=h(y)-h(x)
\label{eq:grev}
\ee
for all $(x,y)\in\cE^2$. In this case, we also have, if the invariant density $\rho_k$ of the driven process is unique, that $\rho_k$ is proportional to $r_k^2\invrho e^{kh}$. This condition on $g$ is verified, in particular, if $g$ is symmetric, $g(x,y)=g(y,x)$.\footnote{This sightly corrects the claim made in \cite{jack2010b} that the driven process is reversible `only if the bias is also time-reversal symmetric: $g(x,y)=g(y,x)$'.} Accordingly, conditioning observables $A_T$, such as the activity, that depend on the jumps of $X_t$, but not on the `direction' of these jumps do not modify the reversibility of $X_t$. The same is true if $A_T$ does not depend on the jumps of the process, that is, if $g=0$ and the conditioning only involves an integral of $f(X_t)$ in time.

These results translate for diffusions as follows. The driven process is reversible with respect to the invariant density $\rho_k$ if and only if its modified drift 
\be
\Fhat_k=\Fhat+D(kg+\nabla\ln r_k),
\ee
obtained from (\ref{eq:tdpe-2}), satisfies
\be
\Fhat_k=\frac{D}{2}\nabla\ln\rho_{k},
\label{eq:diffper}
\ee
which is equivalent to
\be
\Fhat+D\left(kg+\frac{1}{2}\nabla\left(\ln\frac{r_k}{l_k}\right)\right)=0.
\label{eq:cpd}
\ee
This is a functional equation involving $f$ and $g$, via $r_k$ and $l_k$, which is also difficult to solve in general. We can simplify it, as before, by assuming that the reference process $X_t$ is reversible with respect to $\invrho$, as in (\ref{eq:BDD}), in which case
\be
\Fhat_k=\frac{D}{2}\nabla\ln(\invrho r_k^{2})+Dkg.
\label{eq:tdpe-1}
\ee
A particular solution of this equation is obtained if $g$ is gradient, $g=\nabla h/2$. Then the driven diffusion is a reversible diffusion with respect to the invariant density $\rho_k$, which is moreover proportional to $r_k^{2}\invrho e^{kh}$. In particular, if $g=0$ and $D$ is constant, then the driven process is a reversible diffusion with drift given by
\be
F_k=F+D\nabla \ln r_k.
\ee
We thus see for diffusions that conditioning observables $A_T$ that do not depend on the transitions of $X_t$ ($g=0$) or depend on these transitions but via  a gradient perturbation $g$ do not modify the reversibility of $X_t$. 

\subsection{Identities and constraints}

It was found in \cite{baule2008,simha2008,baule2010b} that the driven process admits in many cases certain invariant quantities that constrain its transition rates. These constraints arise very generally and very simply from our results. From (\ref{eq:pesp}), we can write
\begin{eqnarray}
W_k(x,dy)W(y,dx)	&=& r_k(x)^{-2}\, W_{k}(y,dx)W(x,dy)\,r_k(y)^{2}\, e^{k[g(x,y)-g(y,x)]}\nonumber\\
W_k(x,dy)W_k(y,dx)	&=& W(x,dy)W(y,dx)\, e^{k[g(x,y)+g(y,x)]},
\label{eq:cpsp}
\end{eqnarray}
which are the most general identities that can be obtained for the transition rates of the driven process. If $g$ is a symmetric function, they reduce to
\begin{equation}
W_{k}(x,dy)W(y,dx)=r_k(x)^{-2}\, W_{k}(y,dx)W(x,dy)\, r_k(y)^{2},
\label{eq:cpspgs}
\end{equation}
whereas if $g$ is antisymmetric or $g=0$, we find
\be
W_k(x,dy)W_k(y,dx)=W(x,dy)W(y,dx)
\label{eq:cpspga}
\ee
for all $(x,y)\in\cE^2$. The latter result is referred to in \cite{baule2008,baule2010b} as the `product constraint'. An example of symmetric observable is the so-called activity, which is proportional to the number of jumps occurring in a jump process, whereas an example of antisymmetric observable is the current, which assigns opposite signs to a jump and its reversal.

These constraints on the transition rates also imply constraints on the escape rate $\lambda=(W1)$ of a jump process. Integrating (\ref{eq:pesp}) with respect to $y$, keeping in mind that $r_k$ is the right eigenfunction of $\cL_{k}$ associated with $\Lambda_k$, we obtain the following relation between the escape rates of the reference process and those of the driven process: 
\be
\lambda_k(x)=\lambda(x)-kf(x)+\Lambda_{k}.
\label{eq:cpspte}
\end{equation}
In the case $f=0$, this yields
\be
\lambda_k(x)=\lambda(x)+\Lambda_{k}
\label{eq:be}
\ee
for all $x\in\cE$, which implies
\be
\lambda_k(x)-\lambda_k(y)=\lambda(x)-\lambda(y)
\label{eq:be2}
\ee
for all $x\in\cE$, a result referred to as the `exit rate constraint' in \cite{baule2008,baule2010b}.

Diffusive analogs of these constraints can be derived from our results. In the case where the covariance matrix $D$ is invertible, (\ref{eq:tdpe}) implies by taking its exterior derivative that
\be
d\left(D^{-1}F_k-D^{-1}F-kg\right)=0,
\label{eq:cpd-2}
\ee
where all vectors are interpreted as $1$-forms. We conclude from this result that the $1$-form associated with $D^{-1}F_k-D^{-1}F-kg$ is closed. In two and three dimensions, this implies
\be
\nabla\times (D^{-1}F_k)=\nabla\times (D^{-1}F)+k\nabla\times g
\label{eq:gauge1}
\ee
and thus
\be
\nabla\times (D^{-1}F_k)=\nabla\times (D^{-1}F)
\label{eq:gauge2}
\ee
if $g$ is gradient. For diffusions on the circle, (\ref{eq:cpd-2}) only implies that $F_k-F-kg$ is the derivative of a periodic function.\footnote{Periodic functions on the circle are not necessarily the derivative of periodic functions: consider, for example, the constant function.}

These results can be interpreted physically as circulation constraints, showing that the non-reversibility of the driven process, measured by the circulation of its drift, is directly related to the non-reversibility of the reference process and the non-gradient character of $g$. This connection with non-reversible dynamics can be emphasized by rewriting (\ref{eq:gauge1}) in terms of the stationary probability current, defined by
\be
J_{\invrho}=\Fhat\invrho-\frac{D}{2}\nabla \invrho,
\ee
or the so-called probability velocity
\be
V_{\invrho}\equiv\frac{J_{\invrho}}{\invrho}=\Fhat-\frac{D}{2}\nabla \ln \invrho.
\ee
Both vector fields are zero for reversible (equilibrium) diffusions. In terms of $V$, we then obtain
\be
\nabla\times(D^{-1}V_{\rho_k})=\nabla\times (D^{-1}V_{\invrho})+k\nabla \times g,
\ee
where $V_{\rho_k}$ is the probability velocity of the driven process with invariant density $\rho_k$. A similar result applies in higher dimensions by replacing the rotational with the exterior derivative.

\section{Applications}
\label{sec:apps}

We study in this section three applications of our results for Brownian motion, the Ornstein--Uhlenbeck process, and the problem of quasi-stationary distributions. The applications are simple: they are there only to illustrate the different steps needed to obtain the driven process and the effect of boundary dynamics. In the case of quasi-stationary distributions, we also want to show how our results recover known results obtained by a different approach. 

For more complex applications involving many-particle dynamics, such as the totally asymmetric exclusion process and the zero-range process, see \cite{jack2010b,popkov2010,popkov2011,harris2013b,jack2014}. Applications for diffusions can be found in \cite{simha2008,knezevic2014}, whereas applications for quantum systems can be found in \cite{garrahan2010,garrahan2011,ates2012,genway2012,hickey2012}. One interesting aspect of many-particle dynamics is that current-type conditionings have the generic effect of producing long-range interactions between particles at the level of the stationary distribution of the driven process \cite{evans2004,jack2010b,popkov2010,popkov2011,jack2014}. 

In future publications, we will discuss in more detail some of the connections mentioned in the introduction, in particular those relating to conditional limit theorems and optimal control theory, in addition to tackling other applications of our results, including the case of diffusions conditioned on occupation measures,\footnote{F. Angeletti, H. Touchette, in preparation, 2014.} which is relevant for studying metastable states and quasi-stationary distributions. We will also study the low-noise (Freidlin--Wentzell) large deviation limit \cite{freidlin1984}, and develop numerical techniques for obtaining the spectral elements used to construct the driven process.

\subsection{Extensive Brownian bridge}

We revisit Example~\ref{par:OU} about the Wiener process conditioned on reaching the point $X_T=aT$. This observable is a particular case of $A_T$ obtained with $f=0$, $g=1$, and $X_0=0$.

From the Gaussian propagator of the Wiener process, we find
\be
\Pr_x\{X_T/T\in [a,a+da]\}\asymp e^{-TI(a)}da,
\ee
with $I(a)=a^2/2$, as well as
\be
\lim_{T\ra\infty} \frac{1}{T}\ln \Ex_x[e^{k X_T}]=\frac{k^2}{2}.
\ee
The latter result is equal to the dominant eigenvalue $\Lambda_k$, as can be verified from the expression of the tilted generator
\be
\cL_k=\frac{1}{2}\left(\frac{d}{dx}+k\right)^2=\frac{1}{2}\frac{d^2}{dx^2}+k\frac{d}{dx}+\frac{k^2}{2},
\ee
obtained from (\ref{eq:gtpd}). In fact, in this case, we have $r_k(x)=l_k(x)=1$. From (\ref{eq:tdpe}), we thus find that the drift of the driven process, equivalent to the conditioned process, is $F_k=k$. To re-express this drift as a function of the conditioning $X_T/T=a$, we use $I'(a)=a=k$ to obtain $F_{k(a)}=a$. This shows that the process equivalent to the Brownian motion conditioned with $X_T=aT$ is the drifted Brownian motion $W_T+at$, as found previously in (\ref{eq:pew}). Equivalence is for all $a\in\reals$, since $I(a)$ is convex. Moreover, since the typical value of $X_T/T$ is $0$, we have $X_t|X_T=0\procequiv X_t$; that is, the Brownian bridge bridged at $T\ra\infty$ is asymptotically equivalent to the Wiener process.

There is a subtlety involved in this calculation, in that $r_k=l_k=1$ are not normalizable. To circumvent this problem, it seems possible to consider the problem on a compact domain of $\cE=\reals$, such as the interval $[-\ell,\ell]$, to obtain a gapped spectrum with normalizable eigenfunctions, and then take the limit $\ell\ra\infty$. This is a common procedure used in physics, for example, in quantum mechanics to deal with the free particle.

\subsection{Ornstein--Uhlenbeck process}

Consider the Ornstein--Uhlenbeck process, defined in (\ref{eq:OU}), with the conditioning observable
\be
A_T=\frac{1}{T}\int_0^T X_t\, dt,
\label{eq:linobs}
\ee
which corresponds to the choice $f(x)=x$ and $g=0$. The spectral elements of this observable are easily found to be $r_k(x)=e^{kx/\gamma}$ and $\Lambda_k=\sigma^2 k^2/(2\gamma^2)$. From the expression of $r_k$ and (\ref{eq:tdpe}), we then find that the effective drift of the driven process is
\be
F_k(x)=-\gamma x+\frac{\sigma^2 k}{\gamma}.
\ee
With the rate function
\be
I(a)=\sup_k\{ka-\Lambda_k\}=\frac{\gamma^2 a^2}{2\sigma^2},
\ee
we then find
\be
F_{k(a)}(x)=-\gamma x+\frac{a}{\gamma}.
\ee
Hence, the conditioning only adds a constant drift to the process, which ensures that $X_T/T \ra a$ as $T\ra\infty$. Naturally, since the typical value of $A_T$ is $0$ in the original Ornstein--Uhlenbeck process, conditioning on $X_T/T=0$ yields the same process with $F_{k=0}(x)=-\gamma x$.

If instead of choosing the linear observable (\ref{eq:linobs}), we choose
\be
A_T=\frac{1}{T}\int_0^T X_t^2\, dt,
\ee
the same steps can be followed to obtain
\be
F_{k(a)}(x)=-\frac{\sigma^2}{2a}x.
\label{eq:quadobs}
\ee
In this case, the conditioning keeps the linear force of the Ornstein--Uhlenbeck process, but changes its friction coefficient to match the variance of the process with the value of $A_T$. 

To close this example, let us revisit the conditioning $A_T=X_T/T=a$, studied in Example~\ref{par:OU}, which corresponds to the choice $f=0$ and $g=1$. We know from our previous discussion of this example that the driven process cannot describe this conditioning because the latter does not affect the `interior' dynamics of the process in the asymptotic limit $T\ra\infty$. Let us see how this arises in our theory. From the exact form of the propagator (\ref{eq:PTOU-2}), we find that
\be
P\{X_T/T\in [a,a+da]\}\asymp e^{-T^2\gamma a^2/\sigma^2},
\label{eq:PTOU-1}
\ee
so that $I(a)=\infty$ if we take the large deviation limit with the scale $T$, as in (\ref{eqldp}). In this case, we can formally take $\Lambda_k=0$ and $r_k(x)=e^{-kx}$ for the spectral elements, which is consistent with $I(a)=\infty$, to obtain $F_{k(a)}(x)=-\gamma x$. This, as we know from Example~\ref{par:OU}, is the correct interior dynamics produced by the conditioning, but it is not the complete dynamics that actually realizes the conditioning. The problem here is that the large deviation is a boundary effect in time -- it can be seen, physically, as a temporal analog of a `condensation' -- which prevents us from exchanging the two limits in (\ref{eq:ldpcan2}). Consequently, though we can formally define the driven process, it is not equivalent to the conditioned process.

\subsection{Quasi-stationary distributions}

A classical problem in the theory of absorbing processes and quasi-stationary distributions is to condition a Markov chain never to escape from some subset of its state space. We want to briefly show in this subsection that the solution of this problem, obtained classically by defining a new Markov chain restricted on the subset of interest \cite{collet2014}, can be recovered from our results (summarized for Markov chains in Appendix~\ref{app:mc}) by taking the limit $k\ra\infty$. 

To define the problem, let $\{X_i\}_{i=0}^\infty$ be a Markov chain with homogeneous transition matrix $M$. For a subset $\cE_1$ of $\cE$, we consider the conditioning event
\be
\cB=\{\tau_1 >N\},
\ee
where $\tau_1$ is the exit time from $\cE_1$ defined by
\be
\tau_1=\inf\{n: X_n\notin \cE_1\},
\ee
assuming $X_0\in\cE_1$. This means that we are conditioning the Markov chain on leaving $\cE_1$ (or on being `killed' outside $\cE_1$) only after the time $N$. 

Within our theory, this conditioning is effected by considering the observable
\be
A_N=\frac{1}{N}\sum_{i=0}^{N-1} \id_{\cE_1}(X_i),
\ee
or its symmetrized version
\be
A'_{N}=\frac{1}{N}\sum_{i=0}^{N-1}\frac{\id_{\cE_1}(X_i)+\id_{\cE_1}(X_{i+1})}{2}
\ee
for which we have $\id_{\cB}=\id_{A_{N+1}=1}=\id_{A'_{N}=1}$. The second observable leads to the tilted matrix
\be
\cM_k(x,y)\equiv M(x,y) \exp\left[ \frac{k}{2} (\id_{\cE_1}(x)+\id_{\cE_1}(y))\right].
\ee
Given the conditioning $A'_{N}=1$, we must then choose $k\in\partial I(1)$, where $I$ is the rate function of $A'_N$. The form of $\partial I(1)$ depends in general on the Markov chain considered; however, since $I$ is defined on $[0,1]$, we always have $\infty\in\partial I(1)$, so the conditioning follows with the limit $k\ra\infty$.

To see that this limit recovers the correct result, define the matrix
\be
M'(x,y)\equiv \lim_{k\ra\infty} e^{-k} \cM_k(x,y).
\ee
Then,
\be
M'(x,y)=M(x,y)\id_{\cE_1}(x)\id_{\cE_1}(y)=
\left\{
\begin{array}{lll}
M(x,y) & & x,y\in \cE_1\\
0 & & \textrm{otherwise}
\end{array}
\right.
\ee
represents the restriction of the Markov chain on $\cE_1$. Denoting by $\lambda'$ the dominant eigenvalue of $M'$ and by $r'$ its associated right eigenvector, we infer
\be
\lambda'=\lim_{k\ra\infty} e^{-k} \Lambda_k,\qquad r'=\lim_{k\ra\infty} r_k,
\ee
where $\Lambda_k$ and $r_k$ are the corresponding elements of $\cM_k$. 

According to our theory, the effective Markov chain resulting from the asymptotic conditioning $A'_{N}=a$ is given by the generalized Doob transform
\be
M_k(x,y)=\frac{1}{\Lambda_k r_k(x) } \cM_k(x,y) r_k(y).
\ee
Taking the limit $k\ra\infty$, we then obtain
\begin{eqnarray}
\lim_{k\ra\infty} M_k = \frac{1}{\lambda' r'} M' r',
\end{eqnarray}
which is the known result characterizing a Markov chain conditioned on eternally staying in $\cE_1$ \cite{collet2014}. In this context, it can be proved that
\be
\ln \lambda'=\lim_{N\ra\infty} \frac{1}{N}\ln \Pr_{x,M,N}\{\tau_1>N\},
\ee
where $X_0=x\in\cE_1$, so that $\lambda'$ represents the \emph{survival rate} at which the chain stays in $\cE_1$, while the left eigenvector
\be
l'(y)=\lim_{N\ra\infty} \Pr_{x,M,N}\{X_N=y|\tau_1>N\}
\ee
represents the quasi-stationary density of the chain as it stays in $\cE_1$. This last result corresponds to our result (\ref{eq:intl}), and is known as the \emph{Yaglom limit} of the process \cite{collet2014}. Taking the distribution at a time $n<N$ before the conditioning, we obtain instead
\be
l'(y)r'(y)=\lim_{n\ra\infty}\lim_{N\ra\infty} \Pr_{x,M,N}\{X_n=y|\tau_1>N\},
\ee
in agreement with (\ref{eq:intl-2}).

For recent surveys on quasi-stationary distributions, see \cite{villemonais2012,collet2014,doorn2013}; for applications in the context of large deviations, see \cite{chen2012,chen2013b}; finally, see Bauer and Cornu \cite{bauer2014} for a study of the effect of quasi-stationary conditioning on cycle affinities of finite-state jump processes.

\appendix

\section{Derivation of the tilted generator}

\subsection{Pure jump processes}
\label{app:psp}

To derive the form of the tilted generator $\cL_k$ in the case of jump processes, we consider the conservative Markov generator $G_k$ defined by
\begin{equation}
(G_{k}h)(x)=\int_{\cE}W(x,dy)\, e^{kg(x,y)}\, [h(y)-h(x)]
\label{eq:a1}
\end{equation}
for all $x\in\cE$. This generator is only a normalization factor away from $\cL_k$, since
\begin{equation}
G_k=\cL_{k}-(\cL_k1),
\end{equation}
which means that $G_k$ and $\cL_k$ differ only in their diagonal elements.

The process described by $G_k$ is a jump process with transition rates $W(x,dy)e^{kg(x,y)}$. The fact that $e^{k g(x,y)}$ is strictly positive implies that the measure $W(x,dy)e^{kg(x,y)}$ and the original measure $W(x,dy)$ are absolutely continuous, which only means in this context that the $G_k$ and $L$ processes have the same set of allowed jumps $x\ra y$. In this case, we can use Girsanov's Theorem, as applied to jump processes \cite{kipnis1999}, to obtain the Radon--Nikodym of the paths measure of the $G_k$ process with respect to the path measure of the $L$ process:
\be
\frac{d\Pr_{G_k,\mu_0,T}}{d\Pr_{L,\mu_0,T}}(\om)=\exp\left(k\sum_{0\leq t\leq T: \Delta X_t\neq0}g(X_{t^{-}},X_{t^{+}})-\int_0^T dt\, [(We^{kg}1)-(W1)](X_t) \right).
\label{eq:a11}
\ee
Combining this result with the Feynman--Kac formula, we then arrive at
\begin{eqnarray}
\frac{d\Pr_{\cL_{k},\mu_0,T}}{d\Pr_{L,\mu_0T}} & = & \frac{d\Pr_{\cL_{k},\mu_0,T}}{d\Pr_{G_k,\mu_0,T}} \frac{d\Pr_{G_k,\mu_0,T}}{d\Pr_{L,\mu_0,T}}\nonumber \\
 & = & \exp\left(k\sum_{\Delta X_t\neq0}g(X_{t^{-}},X_{t^{+}})-\int_0^T dt\, [(We^{kg}1)-(W1)](X_t)+\int_{0}^{T} (\cL_k1)(X_t)\, dt\right),\nonumber
\label{eq:a111}
\end{eqnarray}
which yields (\ref{eq:ptilt'}) given the expression (\ref{eq:gtps}) of $\cL_k$. 

\subsection{Diffusion processes}
\label{app:df}

Given the generator $L$ of (\ref{eq:gpd}), we introduce a new Markov generator
\be
\cL=L+a\cdot \nabla+b,
\label{eq:a2}
\ee
involving the arbitrary vector field $a$ and scalar field $b$ on $\cE$. Combining the Cameron--Martin--Girsanov Theorem and the Feynman--Kac formula \cite{kac1949,stroock1979,revuz1999}, it can be shown that $\cL$ induces, with the initial measure $\mu_0$, a path measure $\Pr_{\cL,\mu_0,T}$, which is absolutely continuous with respect to the path measure $\Pr_{L,\mu_0,T}$, and whose Radon--Nikodym derivative with respect to the latter measure is
\be
\frac{d\Pr_{\cL,\mu_0,T}}{d\Pr_{L,\mu_0,T}}(\om)=e^{R_T(\om)},
\label{eq:a22}
\ee
where
\begin{eqnarray}
R_{T}&=&\int_0^T D^{-1}(X_t)a(X_t)\circ dX_t\nonumber\\
&&\quad+\int_0^T \left(b(X_t)-D^{-1}(X_t)\, a(X_t)\left(\Fhat+\frac{a}{2}\right)(X_t)-\frac{1}{2}\left(\nabla\cdot a\right)(X_t)\right)\, dt.
\label{forGirs}
\end{eqnarray}

This is a generalization of the Cameron--Martin--Girsanov Theorem for non-conservative processes with $b\neq 0$. In the particular case where $L$ is the generator of the Wiener process $W_t$, and there is no $b$ perturbation, we recover the classical result
\be
R_{T}=\int_0^T\left(a(W_t)dW_t-\frac{a(W_t)^2}{2}\, dt\right),
\label{forGirs-1}
\ee
written here in the It\=o convention \cite{rogers2000}. In our case, we obtain the expression of $\cL_k$ from this general result by equating (\ref{eq:a22}) with (\ref{eq:ptilt'}) to obtain $R_T=kT A_T$, which is solved given (\ref{forGirs}) for $a=kDg$ and 
\be
b=kg\cdot \left(\Fhat+k\frac{Dg}{2}\right)+\frac{k}{2}\nabla\cdot\left(Dg\right)+kf.
\label{eq:a222}
\ee
The expression of $\cL_k$ is therefore
\begin{eqnarray}
\cL_{k} & = & L+kDg\nabla+kg\cdot \left(\Fhat+k\frac{Dg}{2}\right)+\frac{k}{2}\nabla\cdot (Dg)+kf\nonumber\\
 & = & \Fhat\cdot \left(\nabla+kg\right)+(\nabla+kg)\frac{D}{2}(\nabla+kg)+kf.
 \label{eq:a2222222}
 \end{eqnarray}

\section{Change of measure for the generalized Doob transform}
\label{app:chdoob}

To prove (\ref{eq:simi}), it is sufficient to show that 
\be
\Ex_{L^{h,f},\mu_0}[C] = \Ex_{L,\mu_0}\left[C\, h^{-1}(X_0)\, e^{-\int_0^T f(X_{t}) dt}\, h(X_{T})\right]
\label{eq:cyl-1}
\ee
for any cylinder function $C$. In this expression, the generators indicate with respect to which measure the expectation is taken. To arrive at this result, note first that 
\be
e^{tL^{h,f}}(x,dy)=h^{-1}(x)\, e^{t(L-f)}(x,dy)\, h(y)
\label{eq:B11}
\ee
for all $(x,y)\in\cE^2$. Next, replace $t$ by $t-s$ and use the Feynman--Kac formula to express the exponential semi-group as an expectation, so that
\be
e^{(t-s)L^{h,f}}(x,dy)=h^{-1}(x)\, \Ex_{x}\left[e^{-\int_{s}^{t}f(X_u)du}\, \delta(X_{t}-y)dy\right]h(y)
\label{eq:simi2p}
\ee
for all $(x,y)\in\cE^2$. Finally, expand (\ref{eq:cyl-1}) and use (\ref{eq:simi2p}) iteratively to obtain
\begin{eqnarray}
\Ex_{L^{h,f},\mu_0}[C] 
& = & \int_{\cE^{n+1}}C(x_0,\ldots,x_{n})\, \mu_0(dx_0)\, h^{-1}(x_0)\, \Ex_{x_0}[e^{-\int_0^{t_1} f(X_t)dt}\delta(X_{t_{1}}-x_{1})dx_{1}]\cdots\nonumber \\
& &\quad \cdots \Ex_{x_{n-1}}[e^{-\int_{t_n}^{T}f(X_t)dt}\delta(X_T-x_n)dx_n]\, h(x_{n}),
 \label{eq:B111}
 \end{eqnarray}
 which, by multiple integration, is equal to (\ref{eq:cyl-1}).
 
\section{Squared field for diffusion processes}
\label{app:pcarc} 

Let $L$ be the generator of the general diffusion defined in (\ref{eq:gpd}). The application of this generator on the product of two arbitrary functions $f$ and $g$ on $\cE$ yields 
\begin{eqnarray}
(Lfg) &=&f\Fhat\cdot \nabla g+g\Fhat\cdot \nabla f+\frac{1}{2}\nabla\left(g D\nabla f+f D\nabla g \right)\nonumber\\
 & = & f\Fhat \cdot \nabla g+g \Fhat \cdot \nabla f+\frac{g}{2}\nabla D(\nabla f)+\nabla f D \nabla g+\frac{f}{2}\nabla D(\nabla g)\nonumber\\
 & = & f(Lg)+(Lf)g+\nabla f D \nabla g.
\end{eqnarray}
Comparing with the definition (\ref{eq:carrechamp}) of the squared field $\Gamma(f,g)$, we find
\be
\Gamma(f,g)=\nabla f D \nabla g.
\label{eq:C1}
\ee
Putting this result into (\ref{eq:simicar}) with $f=\ln h$, we then find (\ref{adddrift}), which represents the generator of a diffusion process with the same noise fields $\sigma_\alpha$ as the diffusion described by $L$, but with the modified drift given in (\ref{eq:tdpe-2}).

\section{Generator of the canonical path measure}
\label{app:gcpm}

We derive here the time-dependent generator associated with the canonical path measure. To this end, we consider this measure on the cylinder events $\{X_0=x_0,X_{t_1}=x_1,\ldots,X_{t_n}=x_n\}$ with $0\leq t_1\leq \cdots \leq t_n\leq T$ to obtain, following (\ref{eq:ptilt}),
\be
d\Pr_{k,\mu_0,T}^\cano(x_0,\ldots,x_n)=\frac{\mu_0(dx_0)\, e^{t_1 \cL_k}(x_0,dx_1)\, \cdots e^{(t_n-t_{n-1})\cL_k}(x_{n-1},dx_n)\, (e^{(T-t_n)\cL_k}1)(x_n)}{\Ex_{\mu_0}[e^{TkA_T}]},
\ee
with the normalization added according to (\ref{eq:cano}). Therefore,
\be
d\Pr_{k,\mu_0,T}^\cano(x_n|x_0,\ldots,x_{n-1})
\equiv\frac{d\Pr_{k,\mu_0,T}^\cano(x_0,\ldots,x_n)}{d\Pr_{k,\mu_0,T}^\cano(x_0,\ldots,x_{n-1})}
=\frac{e^{(t_n-t_{n-1})\cL_k}(x_{n-1},dx_n)\, (e^{(T-t_n)\cL_k}1)(x_n)}{(e^{(T-t_{n-1})\cL_k}1)(x_{n-1})}
\label{eq:canojoint}
\ee
which shows that the conditional measure $d\Pr_{k,\mu_0,T}^\cano(x_n|x_0,\ldots,x_{n-1})$ is Markovian, since it does not depend on all the previous points $x_0,\ldots,x_{n-1}$ but only on $x_{n-1}$. However, it is non-homogeneous, since it explicitly depends on $t_{n-1}$, $t_n$, and $T$.

To derive the generator $L_{k,t,T}^\cano$ of this Markovian measure, we introduce the positive function
\be
h_{t,T}(x)=(e^{(T-t)\cL_k}1)(x)
\ee
to write the transition probability associated with (\ref{eq:canojoint}) as
\be
P^{\cano,t}_{k,s,T}(x,dy)=\frac{1}{h_{s,T}(x)}e^{(t-s)\cL_k}(x,dy)\, h_{t,T}(y).
\ee
Noting that $h_{t,T}$ solves the backward differential equation 
\be
(\partial_t+\cL_k)h_{t,T}=0,\quad h_{T,T}=1,
\ee
we then have that $h_{t,T}$ is space--time harmonic with respect to $\cL_k$, which implies from (\ref{eq:sth})-(\ref{eq:tdd}) that the canonical measure is the Doob transform of $\cL_k$ with the function $h_{t,T}$ involving $\cL_k$ itself. This means explicitly that
\be
L_{k,t,T}^\cano=(\cL_k)^{(\exp((T-t)\cL_k)1)}.
\ee
This result is valid for $t<T$, but also for $t=T$ which yields
\be
L_{k,T,T}^\cano=\cL_k - (\cL_k1).
\ee 
In the limit $T\ra\infty$, $L_{k,t,T}^\cano$ becomes homogeneous; however, the limit is different for $t<T$ and $t=T$, as shown in (\ref{eq:bg1}) and (\ref{eq:bg2}).

\section{Markov chains}
\label{app:mc}

We briefly re-express in this last section our main results for the simpler case of homogeneous Markov chains. In this context, the generalized Doob transform seems to have appeared for the first time in the work of Miller \cite{miller1961}.

The sequence $X_0,X_1,\ldots, X_{N}$ of random variables is a homogeneous Markov chain if its joint measure is given by
\be
d\Pr_{M,\mu_0,N}(x_1,\ldots, x_{N})=\mu_0(dx_0)M(x_0,dx_{1})....M(x_{N-1},dx_{N}),
\label{eq:D1}
\ee
where $M(x,dy)$ is the transition matrix and $\mu_0$ is the initial measure for $X_0$. The generalized Doob transform of the Markov chain is defined as
\be
M^h(x,dy)=\frac{1}{(M h)(x)} M(x,dy)h(y),
\label{eq:D2}
\ee
where $h$ is a strictly positive function on $\cE$. This transformed matrix remains a stochastic matrix, as shown in \cite{miller1961}, which can be used to define a discrete-time path measure $d\Pr_{M^h,\mu_0,N}$, whose Radon--Nikodym derivative with respect to the original Markov chain is
\be
\frac{d\Pr_{M^h,\mu_0,N}}{d\Pr_{M,\mu_0,N}}(x_0,\ldots,x_{N})=\frac{1}{(Mh)(x_0)}\exp\left(\sum_{i=0}^{N-1}\ln\frac{h(x_i)}{(Mh)(x_i)}\right)h(x_N).
\label{eq:D3}
\ee
This follows by re-expressing in discrete time the proof presented in Appendix~\ref{app:chdoob}.

Consider now the observable
\be
A_N=\frac{1}{N}\sum_{i=0}^{N-1}g(X_{i},X_{i+1}),
\label{eq:osp-1-1}
\ee
where $g:\cE^2\ra\reals$. The tilted generator $\cL_k$ is replaced for this observable by the \emph{tilted matrix}
\be
\cM_k(x,dy)=M(x,dy)e^{kg(x,y)}.
\label{eq:D4}
\ee
The particular observable
\be
A_N=\frac{1}{N}\sum_{i=0}^{N-1}f(X_i)
\ee
is covered by this result simply by taking $g(x,y)=f(x)$, so that we do not have to consider additive and two-point observables separately.

The dominant (Perron--Frobenius) eigenvalue of $\cM_k$ is denoted by $\zeta_k$ and leads to the following result for the SCGF:
\be
\lim_{N\ra\infty}\frac{1}{N}\ln \Ex_{\mu_0}[e^{kNA_N}]=\ln \zeta_k \equiv \Lambda_k.
\ee
Denoting by $r_k$ the right Perron--Frobenius eigenvector of $\cM_k$, we can show, similarly to our previous results, that the Markov chain $\{X_i\}$ conditioned on $A_N=a$ is asymptotically equivalent to a Markov chain described by the following transition matrix:
\be
M_k (x,dy)= \cM_k^{r_k}(x,dy)= \frac{1}{\zeta_k r_k(x)} \cM_k(x,dy) r_k(y).
 \ee
The stationary density of this driven process is the same as in the continuous-time case, namely, $\rho_k(x)=l_k(x)r_k(x)$, where $l_k(x)$ is the left eigenvector of $\cM_k$ associated with $\zeta_k$. Moreover, all our results about the reversibility of this density apply with minor changes.

\begin{acknowledgments}
We would like to thank Patrick Cattiaux, Mike Evans, Rosemary J. Harris, and Vivien Lecomte for useful discussions on this paper. We are also grateful for the hospitality and support of the Laboratoire J.~A. Dieudonn\'e, the Kavli Institute for Theoretical Physics, and the Galileo Galilei Institute for Theoretical Physics, where parts of this work were developed and written. Further financial support was received from the ANR STOSYMAP (ANR-2011-BS01-015), the National Research Foundation of South Africa (CSUR 13090934303) and Stellenbosch University (project funding for new appointee).
\end{acknowledgments}

\bibliography{masterbib}

\end{document}